\documentclass[11pt,a4paper,oneside]{article}
\pdfoutput=1
\usepackage{jheppub}
\usepackage{hyperref}
\usepackage{graphicx,xcolor}
\usepackage{sectsty,amssymb,amsmath}
\usepackage{booktabs}
\usepackage[utf8]{inputenc} 
\usepackage{subcaption}
\usepackage{siunitx}
\usepackage{cleveref}
\usepackage{placeins}
\usepackage{comment}
\usepackage{mathtools}
\allowdisplaybreaks 

\graphicspath{{plots}}

\newcommand{\abs}[1]{\left\lvert#1\right\rvert}

\begin{document}

\title{
A Practical Guide to Sensitivity Projections for Beyond the Standard Model Physics}

\author[a]{Francesca Acanfora,}  
\author[a]{Michael J.~Baker,}   
\author[a]{Joaquim Iguaz Juan,}  
\author[a]{Aidan Symons,}
\author[a]{Andrea Thamm}     

\emailAdd{facanfora@umass.edu} 
\emailAdd{mjbaker@umass.edu}  
\emailAdd{jiguazjuan@umass.edu}
\emailAdd{asymons@umass.edu} 
\emailAdd{athamm@umass.edu}

\affiliation[a]{Department of Physics, University of Massachusetts Amherst, MA 01003, USA}

\date{\today}


\abstract{
Statistics is a well-established discipline, yet its application can be confusing, particularly in fields where the language of statistics is unfamiliar and when making projections, where real data is not yet available.  Both of these factors are commonplace in Beyond the Standard Model physics. In this work, we clarify the distinction between goodness-of-fit tests, which assess the compatibility of an observation with a single hypothesis, and hypothesis tests, which quantify how strongly some data prefer one hypothesis over another. We emphasise that for Beyond the Standard Model sensitivity projections, hypothesis testing should be used. We give detailed examples using both Poisson and multinomially distributed data, deriving exact $p$-values and discussing Monte Carlo and asymptotic approximations. We further demonstrate that for the Asimov data set all the test statistics discussed, for both goodness-of-fit tests and hypothesis tests, can be approximated by Pearson's $\chi^2$ statistic, and the difference in approaches comes down to the distribution of the test statistics.  We also compare goodness-of-fit and hypothesis test $p$-values in two realistic examples. We (1) consider the reach of a muon collider on heavy spin-1 resonances and (2) provide updated projections for the potential of current and future gamma ray telescopes to probe new dark sectors via the observation of Hawking radiation from the explosion of a black hole.  Finally, we discuss the CLs method, the significance and systematic errors.}

\maketitle

\section{Introduction}
\label{sec:intro}

A key part of studying the phenomenology of Beyond the Standard Model (BSM) physics is using statistical tests to quantify how much of the parameter space of a given BSM model a future experiment or measurement can probe, often called a sensitivity projection.  This provides a guide which experimentalists can use to optimise their experiments to target the most motivated regions of parameter space and helps compare the reach of complementary experiments, which is essential in a landscape of limited funding and expensive experiments that operate on very long time scales.

When we are interested in what a future experiment can probe, we do not yet have access to any actual data.  However, almost all resources on the use of statistics in high energy physics focus on the case where the physicist has access to data and wishes to measure certain physical quantities (see, e.g., Refs.~\cite{Lyons:1986em,Barlow:1989:GUS,Cowan:1998ji,James:2006zz,Gross:2018okg,ParticleDataGroup:2026mpi,Segura:2024srj,bini-notes} for pedagogical treatments and Refs.~\cite{Moneta:2010pm,Abasov:2025hsy} for software implementations) and assess the error in their measurement (see, e.g., \cite{Conway:2011in,cowan-slides} for a discussion of nuisance parameters and \cite{Barlow:1993dm,Liu:2023bqc} for Monte Carlo sampling uncertainties).  While this is, of course, an important application of statistics in physics, BSM phenomenologists can be at a loss as to how and why to apply statistical tools to their situation.  While many resources do discuss limit setting in the absence of data, this usually assumes the reader has a working knowledge of how to use the tools when they do have data. 

One important distinction that is rarely made is between a goodness-of-fit test\footnote{Goodness-of-fit tests are sometimes called `significance tests' in the literature.} and a hypothesis test.  Historically, goodness-of-fit tests were developed by Ronald Fisher\footnote{Note that Ronald Fisher was both a very influential scientist and founding chairman of the University of Cambridge Eugenics Society \cite{Aylward}.} while hypothesis tests were a competing framework developed by Jerzy Neyman and Egon Pearson.  A goodness-of-fit test asks how likely it is that some given data comes from some underlying hypothesis, while a hypothesis test asks, given two hypotheses, which one the data is more likely to have come from.  These are very different statistical questions, but quickly become confused as the treatment of both involves (or can involve) test statistics, log-likelihood ratios, $\chi^2$ distributions and Pearson's $\chi^2$ test.  Both approaches finally report a $p$-value, which quantifies the probability of observing results as or more extreme than those measured, assuming a null hypothesis is true. While a $p$-value can be computed exactly in some limited cases (see \cite{Resin:2023} for multinomial-distributed data) or estimated using Monte Carlo methods, both exact $p$-values and Monte-Carlo methods can be computationally expensive so practical analyses typically follow a different route. The Neyman-Pearson lemma \cite{Neyman:1933wgr} (roughly) states that the likelihood ratio is the most powerful test statistic. Then, Wilks' theorem \cite{Wilks:1938dza,cowan-note} (again roughly) states that if one hypothesis is a special case of the other (we will see what this means) then the distribution of $-2 \times \log(\text{likelihood ratio})$ asymptotically follows a $\chi^2$ distribution with a certain number of degrees of freedom. Since the $\chi^2$ distributions are tabulated, the $p$-values can be estimated without extensive computations.

For histogram-based analyses in high-energy physics, Ref.~\cite{Baker:1983tu} systematized this framework, and Refs.~\cite{Cowan:1998ji,Cowan:2010js,cowan-note-2,Abasov:2025hsy} provide a comprehensive treatment of asymptotic likelihood methods. 
For confidence interval construction, classical approaches can yield unphysical exclusion limits when data fluctuate downward. Feldman and Cousins \cite{Feldman:1997qc} addressed this by modifying the test statistic to ensure proper coverage properties. The CLs method \cite{Read:2002hq,read-note,mike-evans-slides,read-slides,cowan-slides-2,gross-slides,bini-slides,sonneveld-slides} takes a complementary approach, incorporating experimental power to prevent overestimating the sensitivity. Both have become standard at the LHC \cite{ATLAS:2011tau, cranmer-slides} and other major experiments, with CLs now preferred for limit-setting in BSM searches.

Sensitivity projections face a distinct challenge: what should we do when there is no observed data.  Ref.~\cite{Cowan:2010js} suggested that we should  assume the future data matches the predictions of the Standard Model (SM) hypothesis (or whichever background model is being assumed). This is known as the Asimov data set, and it enables computation of expected limits under the assumption of no signal.  None-the-less, the scarcity of resources addressing BSM-specific applications has led to imprecise statistical analyses in the literature. For instance, BSM phenomenologists often perform some kind of `$\chi^2$-test' for sensitivity estimation. This is, as we will show, a vague statement that could refer to either a goodness-of-fit test (with some, possibly even an appropriate, number of degrees of freedom) or a hypothesis test. While a `$\chi^2$-test' conventionally denotes a goodness-of-fit test, hypothesis testing is the more widely used framework for BSM analyses. We will show that hypothesis testing is equally straightforward to implement.  While we refrain from listing existing works in the literature which are incorrect or unclear, we will point to some of our own (peer reviewed) work which took the wrong approach.

In summary, in this work we aim to provide a self-contained, pedagogical guide to the use of statistics for the BSM phenomenologist. Rather than presenting abstract formulations we prioritize concrete examples and show how to generalize them to new contexts. We also attempt to make the steps and reasoning as transparent as possible, and provide an accompanying \texttt{Mathematica} notebook on \href{https://github.com/PhysicsBaker/Practical-Guide-to-Sensitivity-Projections}{github} to facilitate reproduction and extension of our results. Note that while Bayesian inference is a useful framework, we here follow a frequentist approach, as is typical in the field.

We organize this paper as follows. \Cref{sec:summary} is an executive summary.  \Cref{sec:Poisson,sec:multinomial} treat Poisson and multinomial distributions, respectively. In both sections, after establishing the statistical foundations, we introduce concrete examples for each distribution and explain how to compute exact $p$-values for both goodness-of-fit tests and hypothesis tests (in \cref{sec:exact-p-value} and \cref{sec:exact-p-value-m}). Since the exact computation of $p$-values is often infeasible, we discuss Monte Carlo and asymptotic approximations (in \cref{sec:approximate-p-value} and \cref{sec:approximate-p-value-m}). 
A central theme of our paper is to try to clarify the crucial distinction between goodness-of-fit tests and hypothesis tests which, though sometimes conflated, answer fundamentally different questions and should not be directly compared. We illustrate this distinction explicitly using realistic BSM examples. In \cref{sec:realistic-example} we study heavy spin-one resonances at a muon collider and in \cref{sec:realistic-example-m} we provide updated projections for the potential of current and future gamma-ray telescopes to probe new dark sectors via the observation of Hawking radiation from the explosion of a black hole. We advocate for hypothesis testing over goodness-of-fit testing in BSM phenomenology, a practice that is already standard in experimental particle physics.  Finally, in \cref{sec:cls} we discuss the CLs method, which is widely used in experimental particle physics, in \cref{sec:significance} we discuss the relationship between $p$-values, significance and $\text{signal}/\sqrt{\text{background}}$, and in \cref{sec:systematic-errors} we discuss systematic errors.

\newpage
\section{Executive Summary}
\label{sec:summary}

Here we give a summary of the practical conclusions of this work.  These are the steps that should be followed to make a sensitivity projection using a hypothesis test.

\begin{enumerate}
    \item {\bf Define the data sets:} First, for your background model, typically the Standard Model, you should compute and bin the expected number of counts of your observable (e.g., the invariant mass of two final state particles, in a collider experiment) over $n_B$ bins. The number of counts in each bin is your SM hypothesis, $H_\text{SM}$ (which is an $n_B$-component vector).  Second, you should do the same for each point in parameter space of your BSM model (where here we mean signal plus background, not signal only).  This will give you the binned BSM hypothesis, $H_\text{BSM}$, for each point in BSM parameter space. Third, to make a sensitivity projection in the absence of any data, we typically assume that the observation coincides with the Standard Model hypothesis. This is called the Asimov data set, $A = H_\text{SM}$.

     \item {\bf Compute the test statistic of the Asimov data set:}  
     If there are more than 10 events in each bin and the difference between the hypotheses is small ($|H_\text{SM}^i-H_\text{BSM}^i| \ll H_\text{BSM}^i$ for all bins) the test statistic of the Asimov data set is approximately
\begin{align}
\label{eq:es-chi-squared}
\Aboxed{
    q_\text{A} \approx \chi^2_\text{A}
    =
    \sum_{i = 1}^{n_B}
    \frac{(H_\text{SM}^i-H_\text{BSM}^i)^2}{H_\text{BSM}^i + \left(\sigma_\text{sys}^i \right)^2}
    \,,
}
\end{align}
where $\sigma_\text{sys}^i$ are uncorrelated systematic errors that you may want to include (see \cref{sec:systematic-errors} for more details on systematic errors). Compute $q_A$ for each point in the BSM parameter space.
If there are fewer than 10 events in a bin, you may want to compute $q_\text{A}$ using, e.g., \cref{eq:qa-asimov-ht-wilks} for Poisson distributed data (which is the typical distribution at counting experiments like a particle collider) or \cref{eq:multi-hypothesis-test-fc-test-statistic} for multinomially distributed data (which is used when the total number of events is fixed).  

\item {\bf Interpret the $q_A$-value:} 
The BSM parameter points with
\begin{align}
\Aboxed{
    q_A \approx \chi^2 \geq 2.71
    \,
}
\end{align}
 are then excluded at the \textbf{95\% confidence level}.  Tracing this line across the BSM parameter space will lead to a sensitivity projection.
For a \textbf{95\% CLs} exclusion (which is typically used by the experimental collider community) you can plot the region with 
\begin{align}
\Aboxed{
    q_A \approx \chi^2 \geq 3.84
    \,.
    }
\end{align}

What happens behind the scenes is explained in detail in \cref{sec:Poisson} and \cref{sec:multinomial}. In short, the $p$-value can be computed for each point in the BSM parameter space by integrating a $\frac{1}{2}\chi^2$ distribution with one degree of freedom (for a hypothesis test) from $q_A$ to $\infty$. A $p$-value smaller than $0.05$ is then directly related to the specific value of the integration limit, $q_A$.

\end{enumerate}

\section{Poisson Distributed Data}
\label{sec:Poisson}

\begin{figure}
\centering
\includegraphics[width=0.5\linewidth]{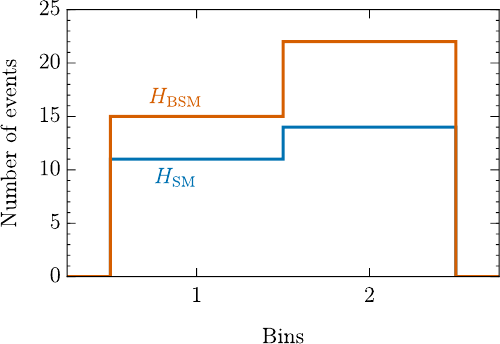}
\caption{The SM and BSM predictions, $H_\text{SM}$ (blue) and $H_\text{BSM}$ (orange), for our two bin Poisson distributed example.}
\label{fig:2-bin-poisson-data}
\end{figure}

To be as concrete as possible we will first discuss a very simple example, where the data is Poisson distributed.  We will imagine that we are doing a counting experiment, like at a collider, and that we have two bins, which could be the invariant mass or pseudorapidity distribution of a two particle final state.  We imagine that, with some given luminosity, kinematic cuts, etc (this is what we will call \textit{an experiment}), the Standard Model predicts that the bins will have
\begin{align}
    H_\text{SM} = (11,14)
\end{align}
events while the Standard Model plus some BSM physics, at a particular point in the BSM parameter space, predicts
\begin{align}
    H_\text{BSM} = (15,22)
\end{align}
events. We plot these distributions in \cref{fig:2-bin-poisson-data}. The SM and BSM expectations are the hypotheses we will be testing.  To compare to language in the literature, when making projections for the sensitivity of a future experiment the
\begin{itemize}
\item \textbf{BSM} hypothesis $H_\text{BSM}$ is the \textbf{null hypothesis} and the
\item \textbf{SM} hypothesis $H_\text{SM}$ is the \textbf{alternate hypothesis}.
\end{itemize}
These can be confused, since when we have experimental data the SM hypothesis plays the role of the null hypothesis~\cite{Cowan:2010js}. In the absence of data, we assume that the observed data coincides with the SM expectation.  The 
\begin{itemize}
\item \textbf{Asimov} data set $A = H_\text{SM}$ is the assumed observed data.
\end{itemize}
In the case that $H_\text{SM}$ does not predict an integer number of events, the Asimov data set is also taken to be non-integer, which does not cause any problems with Poisson distributed data~\cite{Cowan:2010js}.

\subsection{Exact $p$-value}
\label{sec:exact-p-value}

In this section we will compute exact $p$-values for the above two-bin toy example with Poisson distributed data.  We will discuss the computation of the exact $p$-value and show how a goodness-of-fit test differs from a hypothesis test.

In a counting experiment where the statistical error dominates, the number of events observed in the bins will be Poisson distributed and the bins will be independent of each other.  That is, the probability of observing $o_1 = 12$ events in bin 1, if the BSM model is the true model of nature, is
\begin{align}
    P_P^1 &= \frac{(H_\text{BSM}^1)^{o_1} e^{-H_\text{BSM}^1}}{o_1!}
    \\
    &= \frac{15^{12} e^{-15}}{12!}
    \approx 0.08
    \,,
\end{align}
and the probability of observing $o_1 = 12$ events in bin 1 and $o_2 = 20$ events in bin 2 is
\begin{align}
    P_P^{1,2} &= 
    \frac{(H_\text{BSM}^1)^{o_1} e^{-H_\text{BSM}^1}}{o_1!}
    \frac{(H_\text{BSM}^2)^{o_2} e^{-H_\text{BSM}^2}}{o_2!}
    \\
    &= 
    \frac{15^{12} e^{-15}}{12!}
    \frac{22^{20} e^{-22}}{20!}
    \approx 0.007
    \,.
\end{align}
We can now construct the probability mass function, which gives the probability that a discrete random variable is exactly equal to a specific value.
The probability mass function for a BSM hypothesis which predicts $H_\text{BSM}^i$ events in $n_B$ bins and an observation consisting of $o_i$ events is
\begin{align}
\label{eq:poisson-pmf}
    P_P (o|H_\text{BSM})
    &=
    \prod_{i=1}^{n_B}
    \frac{(H_\text{BSM}^i)^{o_i} e^{-H_\text{BSM}^i}}{o_i!}
    \,.
\end{align}
In the literature, this quantity is typically called the \textbf{likelihood}.

In the absence of any data we can now ask: if $H_{\text{BSM}}$ were the true model of nature, how likely would it be to observe data consistent with $H_{\text{SM}}$? The idealized assumption that the observed data is equal to the SM expectation is the Asimov data set, which we denote by
\begin{align}
    A = H_\text{SM}
    \,.
\end{align}
This assumption is made in both goodness-of-fit tests and hypothesis tests.

\subsubsection{Goodness of Fit}

We will now describe a goodness-of-fit test and its associated $p$-value.  Goodness-of-fit tests answer the question: \textit{if $H_\text{BSM}$ were the true model of nature, what is the probability of obtaining an outcome with a probability lower than or equal to that of the observed data by chance alone?} In a goodness-of-fit test there are only two quantities: the hypothesis being tested (here $H_{\text{BSM}}$) and the observation. In the absence of real data we take the observation to be the Asimov data set, $A = H_{\text{SM}}$. The probability of observing events consistent with $A$ if $H_\text{BSM}$ were the true model of nature is
\begin{align}
    P_P^A
    &=
    \prod_{i=1}^{2}
    \frac{(H_\text{BSM}^i)^{o_i} e^{-H_\text{BSM}^i}}{o_i!}
    \\&=
    \prod_{i=1}^{2}
    \frac{(H_\text{BSM}^i)^{A_i} e^{-H_\text{BSM}^i}}{A_i!}
    \\&=
    \frac{15^{11} e^{-15}}{11!}
    \frac{22^{14} e^{-22}}{14!}
    \\&\approx 
    0.0013
    \,.
\end{align}
While this seems unlikely, the most likely outcome, the one that agrees with $H_\text{BSM}$, only has a probability of
\begin{align}
    P_P^\text{BSM}
    &=
    \prod_{i=1}^{2}
    \frac{(H_\text{BSM}^i)^{o_i} e^{-H_\text{BSM}^i}}{o_i!}
    \\&=
    \prod_{i=1}^{2}
    \frac{(H_\text{BSM}^i)^{H_\text{BSM}^i} e^{-H_\text{BSM}^i}}{H_\text{BSM}^i!}
    \\&=
    \frac{15^{15} e^{-15}}{15!}
    \frac{22^{22} e^{-22}}{22!}
    \\&\approx 
    0.0087
    \,.
\end{align}

\begin{figure}
\centering
\includegraphics[width=0.5\linewidth]{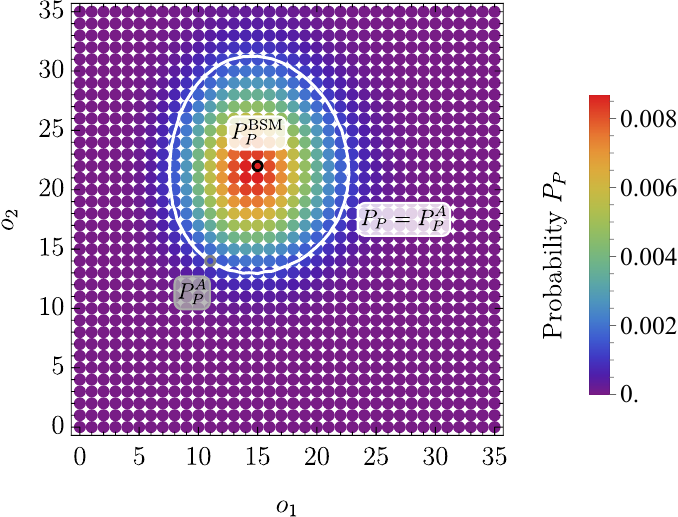}
\caption{The probabilities of observing different counts in the two bins given $H_\text{BSM}$ is the true model of nature.  The white contour corresponds to the probability of obtaining the Asimov data set (equivalent to $H_\text{SM}$).  The positions of $H_\text{BSM}$ and $A$ are given by black and grey circles, respectively.}
\label{fig:exactPValuePoissonGoodnessofFitPlot}
\end{figure}

To better quantify how unlikely it is to get an outcome of $A = H_\text{SM}$ by chance, the {\bf $\mathbf{p}$-value} is used.  The $p$-value is the sum of the probabilities of all possible outcomes with an equal or lower probability than $A = H_\text{SM}$.  The probabilities of observing $o_1$ events in bin 1 and $o_2$ events in bin 2 if $H_\text{BSM}$ were the true model of nature are shown in \cref{fig:exactPValuePoissonGoodnessofFitPlot}.  We see that points near $H_\text{BSM}$ (indicated by the black circle) have the highest probability and the probabilities reduce as we move further away.  The white contour shows where the probabilities are equal to $P_P^A \approx 0.0013$ (with the Asimov data set $A = H_\text{SM}$ highlighted by a grey circle).  

The $p$-value is now given by the sum of the probabilities of all points on or outside the white contour.  However, this is an infinite sum, so instead we can simply compute the sum of probabilities of points within the white circle and subtract it from 1,
\begin{align}
\label{eq:exact-p-value-poisson-goodness-of-fit}
    p &=
    \sum_{\substack{o_1,o_2 \text{ with } \\ P_P(o_1,o_2) \leq P_P^A}} P_P(o_1,o_2)
    \\
    &=
    1-\sum_{\substack{ o_1,o_2 \text{ with } \\ P_P(o_1,o_2) > P_P^A}} P_P(o_1,o_2)
    \\
    &=
    0.15
    \,.
\end{align}
We find a $p$-value of $0.15$. To answer the initial question: if $H_{\text{BSM}}$ were the true model of nature, there is a $15\%$ probability to obtain data as or less compatible with $H_{\text{BSM}}$ than the assumed observed outcome, $A = H_\text{SM}$.
To draw a conclusion, we must choose a \textbf{significance threshold}, $\alpha$, which quantifies when we choose to say some data and a hypothesis are incompatible.   Typically, $\alpha = 0.05$ is chosen. In our example, $p = 0.15 > 0.05$, so the result does not pass the significance threshold and $H_\text{SM}$ could be observed by chance alone if $H_\text{BSM}$ were the true model of nature, at the level of certainty we are willing to accept.

The same result can be phrased in terms of the \textbf{confidence level} (CL), defined as $\mathrm{CL}=1-p$. In our example, $p=0.15$ corresponds to $\mathrm{CL}=85\%$. The commonly used significance threshold $\alpha=0.05$ is equivalent to $\mathrm{CL}=95\%$, which is sometimes adopted when quoting projected exclusion reaches in BSM parameter space. To reach CL$\,= 95\%$ in our example, the white circle would need to be larger, i.e., $H_\text{SM}$ would need to differ more from $H_\text{BSM}$ than it does here. Thus, in this goodness-of-fit test, a future observation consistent with $H_{\text{SM}}$ would \emph{not} exclude $H_{\text{BSM}}$ at the 95\% confidence level.

\subsubsection{Hypothesis Test}

We note that $P_P(o_1,o_2)$ plays two distinct roles in \cref{eq:exact-p-value-poisson-goodness-of-fit}: as the probabilities themselves, and as a condition for which points to include in the sum.  When telling us which points to include in the sum, the probabilities are acting as a test statistic. A {\bf test statistic} is any quantity designed to capture features of the hypotheses that help distinguish the null from the alternative hypothesis. There are many different test statistics and although the probabilities themselves are one possibility, there are many other choices.

Another popular choice is motivated by the Neyman-Pearson lemma, which states (roughly) that for distinguishing two given hypotheses, the most powerful test statistic is the likelihood ratio of the two hypotheses. This is important for hypothesis tests, which answer a fundamentally different question to a goodness-of-fit test.  It is asking: \textit{given two hypotheses, $H_\text{SM}$ and $H_\text{BSM}$, how likely is it that some observed data came from one hypothesis instead of the other}?  The crucial difference to goodness-of-fit tests is that we are now dealing with three quantities: the observation, the SM hypothesis and the BSM hypothesis. We will see that comparing two specific hypotheses is built into the test, independently of what is actually observed.

The likelihood ratio itself is of limited practical use, since it has no simple asymptotic form \cite{vanderVaart1998}. Its logarithm, by contrast, does have a simple asymptotic form when one hypothesis is a special case of the other (we give an example of this condition in \cref{subsec:P-MC-hypothesis-test}).  Wilks' theorem (roughly) says that when one hypothesis is a special case of the other, then in the limit of a large number of events in each bin, the log of the likelihood ratio asymptotically follows a chi squared distribution.  These asymptotic results are important when evaluation of the exact likelihood is too computationally expensive.

Note that the choice between the likelihood ratio and the log of the likelihood ratio makes no difference for the ordering of the outcomes, since $x$ and $\log(x)$ are both monotonically increasing functions. The conditions $P_P \leq P_P^A$ and $\log P_P \leq \log P_P^A$ select the same set of outcome possibilities to include in our sum, so either may be used in the sum in \cref{eq:exact-p-value-poisson-goodness-of-fit}.  

If we want to compare $H_\text{BSM}$ against $H_\text{SM}$ for $o_1$ events counted in bin $1$, the $-2\times$log likelihood ratio is
\begin{align}
    q_1
    &=
    -2 \log\left(
    \frac
    {P_P(o_1|H_\text{BSM})}
    {P_P(o_1|H_\text{SM})}
    \right)
    \\
    &=
    -2 \log\left(
    \frac{(H_\text{BSM}^1)^{o_1} e^{-H_\text{BSM}^1}}{o_1!}
    \cdot\frac{o_1!}{(H_\text{SM}^1)^{o_1} e^{-H_\text{SM}^1}}
    \right)
    \\
    &=
    -2 \left(o_1\log
    \left(
    \frac{H_\text{BSM}^1}
    {H_\text{SM}^1}
    \right)
    -H_\text{BSM}^1
    +H_\text{SM}^1
    \right) \,,
\end{align}
where $P_P(o_1|H)$ is the probability of observing $o_1$ events if $H$ is the true model of nature.  The log likelihood ratio for observing $o_1$ events in bin 1 and $o_2$ events in bin 2 is
\begin{align} \label{eq:test-statistic-P-log-likelihood}
    q
    &=
    -2 \log\left(
    \frac
    {P_P(o|H_\text{BSM})}
    {P_P(o|H_\text{SM})}
    \right)
    \\
    &=
    -2 \sum_{i=1}^2\left(o_i\log
    \left(
    \frac{H_\text{BSM}^i}
    {H_\text{SM}^i}
    \right)
    -H_\text{BSM}^i
    +H_\text{SM}^i
    \right)
    \,,
\end{align}
where we used the likelihood in \cref{eq:poisson-pmf} in the second step.
This particular choice of test statistic is the {\bf log likelihood ratio} (and the factor $-2$ is included for later convenience). We see that this test statistic contains information about $H_\text{SM}$ even for data that do not agree with $H_\text{SM}$, in contrast to the probability in \cref{eq:poisson-pmf} (which is also the test statistic for a goodness-of-fit test).

\begin{figure}
\centering
\includegraphics[width=0.49\linewidth]{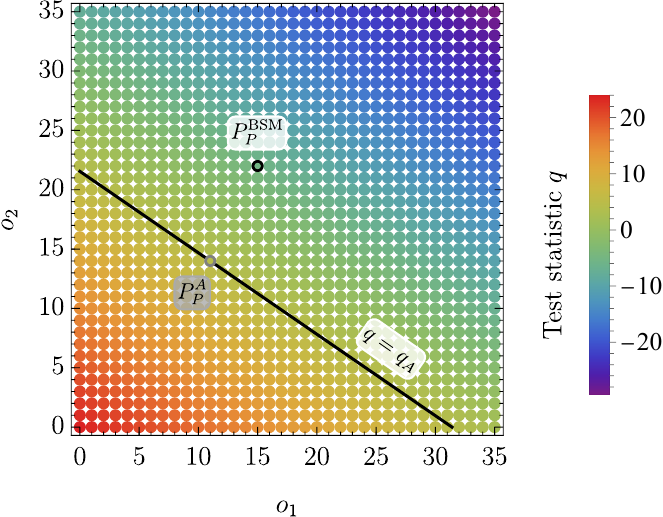}
\includegraphics[width=0.49\linewidth]{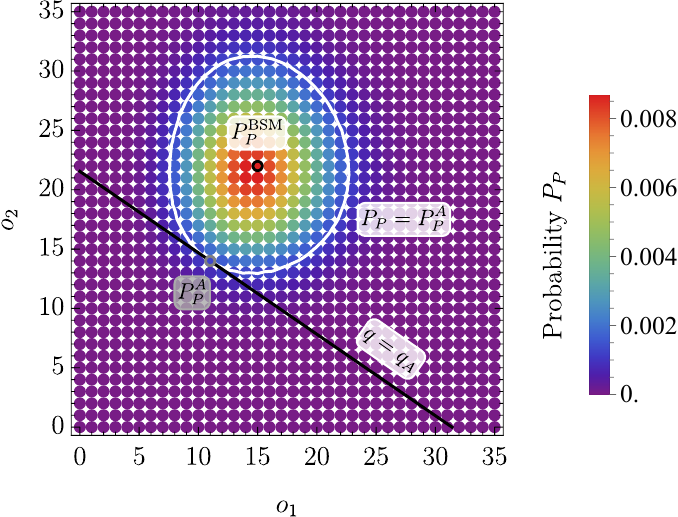}
\caption{Left: The test statistic $q$ for different counts in the two bins given $H_\text{BSM}$ is the true model of nature.  Right: The corresponding probabilities with the contour $P_P = P_P^A$ in white.}
\label{fig:exactPValuePoissonHypothesisTestPlot}
\end{figure}

The test statistic for data consistent with $H_\text{SM} = (11,14)$, the Asimov data set, is
\begin{align}
    q_\text{A} 
    &=
    -2 \sum_{i=1}^2\left(A^i\log
    \left(
    \frac{H_\text{BSM}^i}
    {H_\text{SM}^i}
    \right)
    -H_\text{BSM}^i
    +H_\text{SM}^i
    \right)
    \\
    &=4.5
    \,.
\end{align}
The test statistic for data consistent with $H_\text{BSM} = (15,22)$ is $q_\text{BSM} = -5.2$.  The test statistic for $o_1 = 12$ events in bin $1$ and $o_2 = 20$ events in bin $2$ is $q = -1.5$.  We see that positive test statistics are more SM like while negative test statistics are more BSM like. We can now compute the test statistic for any possible observation in the two bins, shown in the left panel of \cref{fig:exactPValuePoissonHypothesisTestPlot}. The black line indicates test statistics equivalent to $q_\text{A}$. We see that SM-like observations, with positive test statistics, lie in the lower left while BSM-like observations lie in the upper right. 

Now, to compute the $p$-value for the hypothesis test we add up the probabilities of the possibilities with a test statistic equal to or more extreme than $A$.  That is, those to the lower left of the black line in \cref{fig:exactPValuePoissonHypothesisTestPlot} (left),
\begin{align}
\label{eq:exact-p-value-poisson-hypothesis-test}
    p &=
    \sum_{\substack{o_1,o_2 \text{ with } \\ q(o_1,o_2) \ge q_\text{A}}} P_P(o_1,o_2)
    \\
    &=
    0.019
    \,.
\end{align}
We now see that the hypothesis test gives a much smaller $p$-value than the goodness-of-fit test, and a future observation the same as $A = H_\text{SM}$ would exclude $H_\text{BSM}$ at 95\%.

To connect the log-likelihood–ratio test statistic to the goodness-of-fit test, where the test statistic is equivalent to the probabilities themselves, we also display $P_P(o_1,o_2)$ in \cref{fig:exactPValuePoissonHypothesisTestPlot} (right). This makes the distinction between the two procedures explicit: in the goodness-of-fit test we sum over all outcomes outside the white circle, whereas in the hypothesis test we sum only over outcomes to the left of the black line. Because this region is only a subset of the points outside the circle, the hypothesis-test $p$-value is smaller than the goodness-of-fit $p$-value. In this sense, the hypothesis test is more sensitive for discriminating $H_{\text{SM}}$ from $H_{\text{BSM}}$.  In the hypothesis testing scenario, the $p$-value corresponds to $\alpha$, the probability of making a type I error (that is, `rejecting a true null hypothesis' or `rejecting $H_\text{BSM}$ when it is the true model of nature').  In this framework one can also calculate the probability of making a type II error (rejecting a true alternate hypothesis).

Even though goodness-of-fit tests and hypothesis tests are sometimes treated as being interchangeable in the literature, we see that they give quantitatively different results as they answer fundamentally different questions.  In our example, if data is observed which agrees with the SM, the goodness-of-fit test would not exclude the BSM hypothesis, while the hypothesis test would.  We will come back to this point later, but for a consistent comparison between current and future experimental results we advocate the use of hypothesis tests instead of goodness-of-fit tests.

\subsection{Monte Carlo and Asymptotic Estimates of the $p$-value}
\label{sec:approximate-p-value}

In the last section we have worked with a two-bin example with few events where the $p$-values can be computed exactly.  Unfortunately, in most practical settings there are too many bins and too many events to compute exact $p$-values.  In these cases there are a range of approximate methods which can be used.  In this section we show how to approximate $p$-values for Poisson distributed data using Monte Carlo methods and asymptotic methods.  We will first consider the two-bin Poisson distributed scenario discussed above, with 
\begin{align}
    H_\text{SM} = (11,14)
    \,,
\end{align}
and
\begin{align}
    H_\text{BSM} = (15,22)
    \,.
\end{align}

\subsubsection{Goodness of Fit}
\label{subsec:P-MC-goodness}

As in the exact $p$-value computation above, we obtain a goodness-of-fit test when we use probabilities as the test statistic to compare different outcomes.  We will now approximate the $p$-value with a Monte Carlo method.

\begin{figure}
\centering
\includegraphics[width=0.49\linewidth]{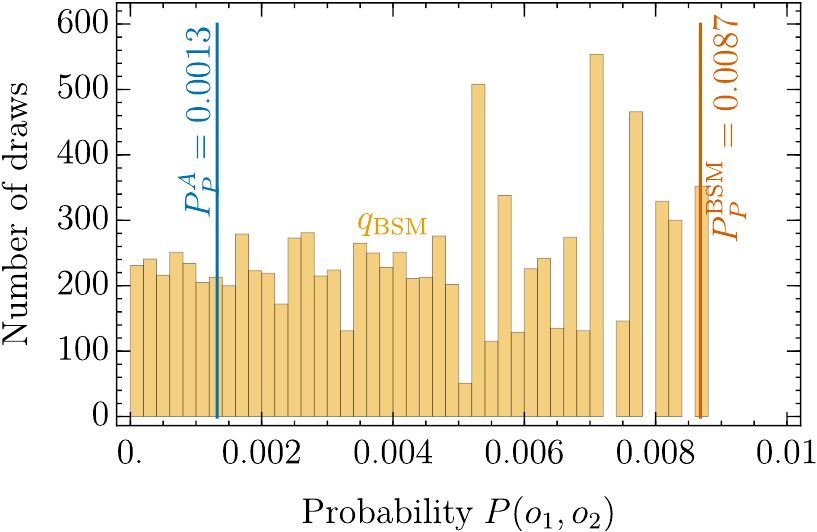}
\includegraphics[width=0.49\linewidth]{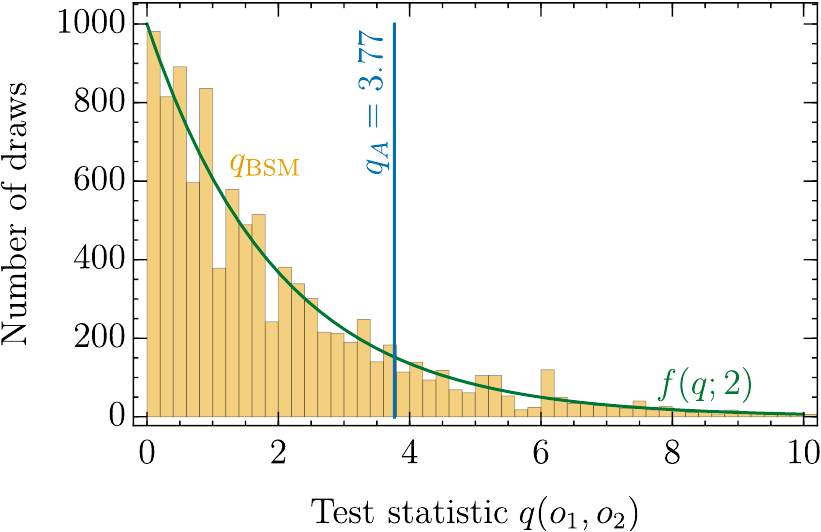}
\caption{Histograms of the probabilities (left) and the test statistic given in \cref{eq:test-statistic-P-goodness-of-fit} (right) obtained from $10^4$ draws from $H_\text{BSM}$, along with the probability and test statistic for the Asimov data set shown in blue.  The orange line shows the probability of observing $H_\text{BSM}$ and the green line shows a $\chi^2$ distribution with 2 degrees of freedom.}
\label{fig:MCPoissonProbsPlot}
\end{figure}

\paragraph{Monte Carlo Methods:}
First we will draw from the BSM hypothesis a large number of times, say $N_\text{MC} = 10^4$.  That is, on each trial we choose outcome $(o_1,o_2)$ with probability $P_P(o_1,o_2)$ where the probability is given by \cref{eq:poisson-pmf}.  A histogram showing the binned probability distribution for an example set of draws is shown in \cref{fig:MCPoissonProbsPlot} (left).  Recall that the most probable outcome, $H_\text{BSM}=(o_1,o_2) = (15,22)$ has probability $P _P^\text{BSM} = 0.0087$ while the probability of obtaining the Asimov data set, i.e., an outcome equal to $H_\text{SM}$, $(o_1,o_2) = (11,14)$, is $P^A_P = 0.0013$.  Note that while the more probable outcomes are drawn more often, there are more outcomes with smaller probabilities.  In the end, the binned distribution of probabilities is approximately uniform although with some large deviations at larger probabilities.  The $p$-value of the goodness-of-fit test can be obtained by counting the number of draws with $P(o_1,o_2) \leq P_P^A = 0.0013$ and dividing by the number of Monte Carlo trials.  The result for this particular set of draws is
\begin{align}
    p &= 
    \frac{1473}{10^4}
    \\
    &\approx
    0.15
    \,,
\end{align}
in agreement with the exact method.

We note that the shape of \cref{fig:MCPoissonProbsPlot} (left) is predominantly due to the underlying distribution, not to the Monte Carlo error.  The various peaks and dips are due to precise features of this particular example.  That means that it is difficult to predict the shape for any given $H_\text{BSM}$ and $H_\text{SM}$.  This can be resolved, in the large number of events limit, by changing the test statistic.  As long as we choose a test statistic that does not change the ordering of the draws, then we can sort using the test statistic and add up the number of draws with a more extreme test statistic than $H_\text{SM}$.  I.e., for draws $a$ and $b$, if $P_P(o_1^a,o_2^a) < P_P(o_1^b,o_2^b)$ then $q(o_1^a,o_2^a) > q(o_1^b,o_2^b)$, where the $<$ changes to a $>$ because of the choice of sign in the test statistic we will use.

If we use the test statistic
\begin{align} 
\label{eq:test-statistic-P-goodness-of-fit}
    q
    &=
    -2 \log\left(
    \frac
    {P(o|H_\text{BSM})}
    {P(H_\text{BSM}|H_\text{BSM})}
    \right)
    \\
    &=
    -2 \log\left(
    \prod_{i=1}^2
    \frac{(H_\text{BSM}^i)^{o_i} e^{-H_\text{BSM}^i}}{o_i!}
    \frac{H_\text{BSM}^i!}{(H_\text{BSM}^i)^{H_\text{BSM}^i} e^{-H_\text{BSM}^i}}
    \right)
    \\
    &=
    -2 \sum_{i=1}^2
    \log
    \left(
    \frac{(H_\text{BSM}^i)^{o_i}}
    {(H_\text{BSM}^i)^{H_\text{BSM}^i}}
    \frac{H_\text{BSM}^i!}{o_i!}
    \right)
    \,,
\end{align}
where the denominator $P(H_\text{BSM}|H_\text{BSM})$ ensures that $q$ is positive (since $P(o|H_\text{BSM}) \leq P(H_\text{BSM}|H_\text{BSM})$ for all possible observations $o$), we find the distribution shown in \cref{fig:MCPoissonProbsPlot} (right).  The value of the test statistic of the Asimov data set is 
\begin{align} 
\label{eq:p-qa-asimov-gof}
    q_\text{A} 
    &=
    -2 \sum_{i=1}^2
    \log
    \left(
    \frac{(H_\text{BSM}^i)^{A^i}}
    {(H_\text{BSM}^i)^{H_\text{BSM}^i}}
    \frac{H_\text{BSM}^i!}{A^i!}
    \right)
    \\
    &= 
    3.77
    \,.
\end{align}
Again to estimate the $p$-value we count the number of draws more extreme than the Asimov data set (in this case to the right of the blue line in \cref{fig:MCPoissonProbsPlot}) and divide by the number of draws, giving
\begin{align}
    p &= 
    \frac{1473}{10^4}
    \\
    &\approx
    0.15
    \,.
\end{align}

Note that this test statistic only involves $H_\text{BSM}$, so is not the log-likelihood ratio used in hypothesis tests.  We have just exchanged one monotonic ordering of the probabilities for another.  However, this particular ordering leads us to the idea of asymptotic approximations to the distributions of test statistics.

\paragraph{Asymptotic Distribution of the Test Statistic:}
We see that the distribution of test statistics in \cref{fig:MCPoissonProbsPlot} (right) is well fitted by the probability density function (p.d.f.) of a $\chi^2$-distribution with $2$ degrees of freedom.  The general form of the $\chi^2$ distribution with $k$ degrees of freedom is
\begin{align} 
\label{eq:chi2-distribution}
    f(q;k)
    &=
    \frac{q^{\frac{k}{2}-1}e^{-\frac{q}{2}}}{2^{\frac{k}{2}} \Gamma\left(\frac{k}{2}\right)}
    \,,
\end{align}
for $q > 0$ where $\Gamma(k/2)$ is the Gamma function.  For example, for $k=2$ we have
\begin{align} 
\label{eq:chi2-distribution-k=2}
    f(q;2)
    &=
    \frac{1}{2}e^{-\frac{q}{2}}
    \,.
\end{align}
In general, for hypotheses with $k$ bins the distribution of test statistics given by \cref{eq:test-statistic-P-goodness-of-fit} is well fitted by the p.d.f.~of a $\chi^2$-distribution with $k$ degrees of freedom~\cite{ParticleDataGroup:2026mpi,ferguson2017course}.  Note that the $-2$ factor in the test statistic in \cref{eq:test-statistic-P-goodness-of-fit} is chosen precisely so that the resulting distribution of test statistics matches the p.d.f.~of a $\chi^2$-distribution.  

Note that the degrees of freedom discussed here are statistical degrees of freedom, they have nothing to do with the parameters of the BSM model.  When making projections of exclusion curves on BSM model parameter space, we are performing statistical tests on each point independently; the number of parameters in the BSM model is unrelated to the number of statistical degrees of freedom.\footnote{This can be confusing when, for example, referring to literature that discusses fitting statistical models to data.  These models may contain degrees of freedom that need to be accounted for.  This is usually not what we are doing in BSM phenomenology.}

We can therefore estimate the $p$-value by integrating $f(q;2)$ from $q_\text{A}$ to $\infty$,
\begin{align} \label{eq:p-value-P-goodness-of-fit-asymptotic}
    p
    &=
    \int_{q_\text{A}}^\infty f(q;2)\,dq
    \\
    &=
    e^{-\frac{q_\text{A}}{2}}
    \\
    &\approx
    0.15
    \,.
\end{align}
This allows us to estimate the $p$-value without doing an exact calculation (which becomes prohibitive for more bins and more event counts) or performing Monte Carlo simulations.

\paragraph{Asymptotic Limit of the Test Statistic of the Asimov Data Set:}
We can then go further and consider the limits where the BSM hypothesis and the Asimov data set do not differ by a large amount in any bin,
\begin{align}
    |\delta^i|
    \ll
    1
    \qquad\text{for all $i$, where}\qquad
    \delta^i
    &=
    \frac{A^i - H_\text{BSM}^i}{H_\text{BSM}^i}
    \,,
\end{align}
and there is a large number of events in each bin,
\begin{align}
    H_\text{BSM}^i
    \gtrsim
    10
\end{align}
for all $i$.\footnote{Note that since we will do two separate expansions, we are also technically assuming that $\delta^i \gg 1/H_\text{BSM}^i$ in all bins.  We also make the same assumption in the following sections when we connect the test statistics of the Asimov data set to Pearson's $\chi^2$ test statistic.}  These approximations are what is meant by an asymptotic estimate of the $p$-value.  The test statistic for the Asimov data set, which coincides with $H_\text{SM}$, is
\begin{align}
    q_\text{A}
    &=
    -2 \sum_{i=1}^2
    \log
    \left(
    \frac{(H_\text{BSM}^i)^{A^i}}
    {(H_\text{BSM}^i)^{H_\text{BSM}^i}}
    \frac{H_\text{BSM}^i!}{A^i!}
    \right)
    \\
    &\approx
    -2 \sum_{i=1}^2
     \bigg(
    A^i
    \log
    \left(
    H_\text{BSM}^i
    \right)
    -
    H_\text{BSM}^i
    \log
    \left(
    H_\text{BSM}^i
    \right)
    \\
    &\hspace{2cm}
    +
    H_\text{BSM}^i
    \log
    \left(H_\text{BSM}^i
    \right)
    -
    H_\text{BSM}^i
    -
    A^i
    \log
    \left(A^i
    \right)
    +
    A^i 
    \bigg)
    \nonumber\\
    &=
    -2 \sum_{i=1}^2
     \bigg(
    H_\text{BSM}^i (1 + \delta^i)
    \log
    \left(
    H_\text{BSM}^i
    \right)
    -
    H_\text{BSM}^i
    \\
    &\hspace{2cm}
    -
    H_\text{BSM}^i (1 + \delta^i)
    \log
    \left(H_\text{BSM}^i (1 + \delta^i)
    \right)
    +
    H_\text{BSM}^i (1 + \delta^i)
    \bigg)
    \nonumber\\
    &=
    -2 \sum_{i=1}^2
     \bigg(
    -
    H_\text{BSM}^i (1 + \delta^i)
    \log
    \left(1 + \delta^i
    \right)
    +
    H_\text{BSM}^i \delta^i
    \bigg)
    \\
    &\approx
    -2 \sum_{i=1}^2
     \bigg(
    -
    H_\text{BSM}^i (1 + \delta^i)
    \left(\delta^i - \frac{(\delta^i)^2}{2}
    \right)
    +
    H_\text{BSM}^i \delta^i
    \bigg)
    \\
    &\approx
    -2 \sum_{i=1}^2
     \bigg(
    -
    H_\text{BSM}^i 
    \left(\delta^i + \frac{(\delta^i)^2}{2}
    \right)
    +
    H_\text{BSM}^i \delta^i
    \bigg)
    \\
    &=
    -2 \sum_{i=1}^2
     \bigg(
    -
    H_\text{BSM}^i 
    \frac{(\delta^i)^2}{2}
    \bigg)
    \\
    &=
    \sum_{i=1}^2
    \frac{(A^i - H_\text{BSM}^i)^2}{H_\text{BSM}^i}
    \,,
\end{align}
where we have used Stirling's approximation $\log(n!) \approx n\log(n) - n$, which is valid for large $n$, and $\log(1+\delta) \approx \delta - \delta^2/2$, which is valid for small $\delta$.  In this situation we can therefore approximate the test statistic of the Asimov data set with \textbf{Pearson's $\chi^2$ test statistic}
\begin{align}
    q_\text{A} \approx \chi^2_\text{A}
    &=
    \sum_{i=1}^2
    \frac{(A^i - H_\text{BSM}^i)^2}{H_\text{BSM}^i}
    \,.
\end{align}
In our two-bin example
\begin{align}
    q_\text{A} \approx \chi^2_\text{A}
    &=
    3.98
    \,,
\end{align}
and using the observation that the distribution of test statistics is approximately $f(q;2)$ we can estimate the $p$-value using \textbf{Pearson's $\chi^2$ test},
\begin{align}
    p
    &=
    \int_{\chi^2_\text{A}}^\infty f(q;2)\,dq
    \\
    &=
    e^{-\frac{\chi^2_\text{A}}{2}}
    \\
    &\approx
    0.14
    \,.
\end{align}
In practice, goodness-of-fit tests usually use Pearson's $\chi^2$ tests, and now we have seen the steps required to show that it approximates the exact $p$-value.  We see that even though we are considering an example with less than 23 events in each bin, and the differences between the hypotheses are quite large (8 events in the case of bin 2), all approximate methods give a good estimate of the exact $p$-value.

\subsubsection{Hypothesis Test}
\label{subsec:P-MC-hypothesis-test}

We now turn to Monte Carlo and asymptotic methods for hypothesis tests of Poisson distributed data.

\begin{figure}
\centering
\includegraphics[width=0.5\linewidth]{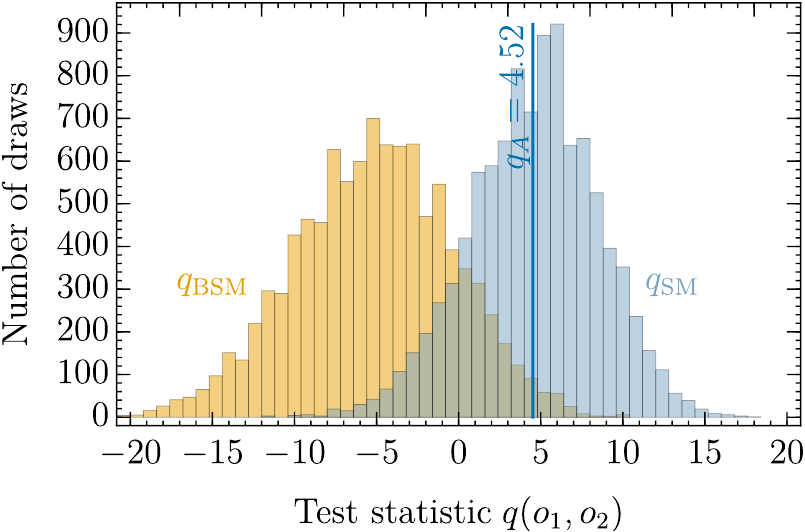}
\caption{Histograms of the test statistics obtained from $10^4$ draws from $H_\text{BSM}$ (orange) and from $H_\text{SM}$ (blue), along with the test statistic for the Asimov data set, $q_\text{A}$.}
\label{fig:MCPoissonTestStatisticPlot}
\end{figure}

\paragraph{Monte Carlo Methods:}
This time we draw from both the SM and BSM hypotheses a large number of times, $N_\text{MC} = 10^4$, using the fact that we expect the observations to be Poisson distributed around the expected numbers of events.  For each draw, $o$, we then compute the log likelihood ratio test statistic as given in \cref{eq:test-statistic-P-log-likelihood},
\begin{align} \label{eq:poisson-hypothesis-test-test-statistic}
    q
    &=
    -2 \log\left(
    \frac
    {P(o|H_\text{BSM})}
    {P(o|H_\text{SM})}
    \right)
    \\
    &=
    -2 \sum_{i=1}^2\left(o_i\log
    \left(
    \frac{H_\text{BSM}^i}
    {H_\text{SM}^i}
    \right)
    -H_\text{BSM}^i
    +H_\text{SM}^i
    \right)
    \,.
\end{align}
The distribution of test statistics for the possible observations expected by chance from the SM and BSM hypotheses are shown in \cref{fig:MCPoissonTestStatisticPlot}.  To find an estimate of the $p$-value we count the proportion of events drawn from $H_\text{BSM}$ with a test statistic more extreme than that of the Asimov data set,
\begin{align}
\label{eq:poisson-hypothesis-test-asimov-test-statistic}
    q_\text{A} 
    &=
    -2 \sum_{i=1}^2\left(A^i\log
    \left(
    \frac{H_\text{BSM}^i}
    {H_\text{SM}^i}
    \right)
    -H_\text{BSM}^i
    +H_\text{SM}^i
    \right)
    \\
    &= 4.52
    \,.
\end{align}
That is, we count how many draws from $H_\text{SM}$ (orange) lie to the right of the blue line in \cref{fig:MCPoissonTestStatisticPlot} and divide by $N_\text{MC}$, giving
\begin{align}
    p &= 
    \frac{205}{10^4}
    \\
    &\approx
    0.02
    \,,
\end{align}
in agreement with the exact $p$-value computed in \cref{eq:exact-p-value-poisson-hypothesis-test}.

\paragraph{Wilks' Theorem:}
As before when we used probabilities to compute the goodness-of-fit $p$-value using Monte Carlo techniques, the above formulation does not readily lead to asymptotic approximations.  We now discuss a method of applying Wilks' theorem, which does lead to an asymptotic approximation (see, e.g., Ref.~\cite{Kahlhoefer:2019vhz} for a short discussion of this method).

\begin{figure}
\centering
\includegraphics[width=0.5\linewidth]{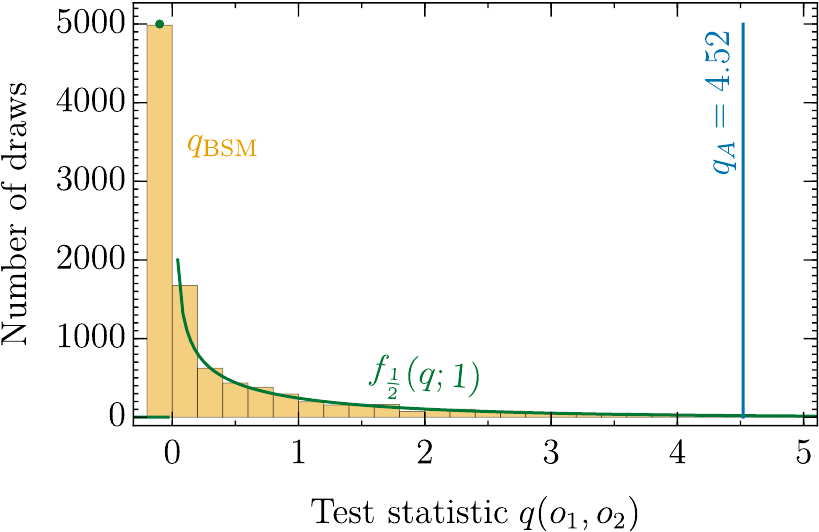}
\caption{Histogram of the test statistics obtained from $10^4$ draws from $H_\text{BSM}$ (orange) using the $H(\mu)$ as the alternate hypothesis, along with the test statistic for the Asimov data set, $q_\text{A}$, in blue.  The green curve and dot illustrate a $\frac{1}{2} \chi^2$ distribution with 1 degree of freedom.}
\label{fig:MCPoissonTestStatisticFCPlot}
\end{figure}

In this method we change the test statistic used above, \cref{eq:poisson-hypothesis-test-test-statistic} and replace the alternate hypothesis $H_\text{SM}$ with a one parameter family of hypotheses,
\begin{align}
    H(\mu) = \mu H_\text{BSM} + (1-\mu) H_\text{SM}
    \,,
\end{align}
which smoothly interpolate between $H_\text{SM}$ and $H_\text{BSM}$ as the parameter $\mu$ changes from 0 to 1.  The null hypothesis $H_\text{BSM}$ then lies in the space of possibilities of the alternate hypothesis $H(\mu)$ (that is, $H_\text{BSM}$ is a special case of $H(\mu)$).  We draw from $H_\text{BSM}$ a large number of times, $N_\text{MC} = 10^4$, and compute the test statistic,
\begin{align} \label{eq:test-statistic-P-hypothesis}
    q
    &=
    \max_{0 \leq \mu \leq 1}
    \left(
    -2 \log\left(
    \frac
    {P(o|H_\text{BSM})}
    {P(o|H(\mu))}
    \right)
    \right)
    \\
    &=
    \max_{0 \leq \mu \leq 1}
    \left(
    -2 \sum_{i=1}^2\left(o_i\log
    \left(
    \frac{H_\text{BSM}^i}
    {H(\mu)^i}
    \right)
    -H_\text{BSM}^i
    +H(\mu)^i
    \right)
    \right)
    \,.
\end{align}
for each draw.  For each draw a different value of $\mu$ is found which maximises the test statistic.  We show the distribution of test statistics in \cref{fig:MCPoissonTestStatisticFCPlot}.

The test statistic for the Asimov data set $o = A = H_\text{SM}$ is then
\begin{align}
    q_\text{A}
    &=
    \max_{0 \leq \mu \leq 1}
    \left(
    -2 \log\left(
    \frac
    {P(A|H_\text{BSM})}
    {P(A|H(\mu))}
    \right)
    \right)
    \\
    &=
    \max_{0 \leq \mu \leq 1}
    \left(
    -2 \sum_{i=1}^2\left(A^i\log
    \left(
    \frac{H_\text{BSM}^i}
    {H(\mu)^i}
    \right)
    -H_\text{BSM}^i
    +H(\mu)^i
    \right)
    \right) \\
    \label{eq:qa-asimov-ht-wilks}
    &=
    -2 \sum_{i=1}^2\left(A^i\log
    \left(
    \frac{H_\text{BSM}^i}
    {A^i}
    \right)
    -H_\text{BSM}^i
    +A^i
    \right)
    \\
    &=
    4.52
    \,,
\end{align}
where we no longer need to maximize since for the Asimov data set the maximum likelihood is given by $\mu = 0$ (since $H(0) = H_\text{SM} = A$) and the test statistic agrees with \cref{eq:poisson-hypothesis-test-asimov-test-statistic}.  We estimate the $p$-value by counting the number of events to the right of $q_\text{A}$ and dividing by the total number of draws.  In this case we find
\begin{align}
    p &= 
    \frac{185}{10^4}
    \\
    &\approx
    0.0185
    \,.
\end{align}
We see that this method provides a good estimate of the exact $p$-value.

\paragraph{Asymptotic Distribution of the Test Statistic:}
We see that the distribution of the test statistic now closely follows a $\frac{1}{2} \chi^2$ distribution with one degree of freedom.  A $\frac{1}{2} \chi^2$ distribution has half of the distribution described by a delta function at $q=0$ while the other half follows a usual $\chi^2$ distribution with one degree of freedom,
\begin{align}
\label{eq:half-chi-squared}
    f_{\frac{1}{2}}(q;1)
    &=
    \frac{1}{2}\delta(q) + \frac{1}{2}f(q;1)
    \,.
\end{align}
Note that there is one degree of freedom because of the one parameter maximum likelihood estimator, parameterised by $\mu$, not because of the number of bins.  If there were more bins but still a one parameter maximum likelihood estimator, then the distribution of test statistics would still asymptote to a $\frac{1}{2} \chi^2$ distribution with one degree of freedom.  We can then analytically estimate the $p$-value using
\begin{align}
\label{eq:test-statistic-P-hypothesis-1-2-chi^2}
    p
    &=
    \int_{q_\text{A}}^\infty f_{\frac{1}{2}}(q;1)\,dq
    \\
    &=
    \int_{q_\text{A}}^\infty \frac{1}{2}f(q;1)\,dq
    \\
    &=
    \int_{q_\text{A}}^\infty \frac{1}{2}
    \frac{e^{-\frac{q}{2}}}{\sqrt{2\pi q}}
    \,dq
    \\
    &=
    \int_{\sqrt{q_\text{A}}}^\infty \frac{1}{\sqrt{2\pi}}
    e^{-\frac{t^2}{2}}
    \,dt
    \\
    &=
    1-
    \frac{1}{\sqrt{2\pi}}
    \int_{-\infty}^{\sqrt{q_\text{A}}} 
    e^{-\frac{t^2}{2}}
    \,dt
    \\
    &=
    1-
    \Phi(\sqrt{q_\text{A}})
    \\
    &=
    \frac{1}{2}
   \text{erfc}\left(\sqrt{\frac{q_\text{A}}{2}}\right)
    \\
    &\approx
    0.017
    \,,
\end{align}
where $\Phi(x)$ is the cumulative distribution function of a normal distribution with mean 0 and standard deviation 1,
\begin{align}
    \Phi(x) = 
    \frac{1}{\sqrt{2\pi}}
    \int_{-\infty}^x
    e^{-\frac{t^2}{2}} \,dt
    \,,
\end{align}
and erfc is the complementary error function.  We see that again this provides another good estimate of the exact $p$-value.

\paragraph{Asymptotic Limit of the Test Statistic of the Asimov Data Set:}

We now consider the limit where the two hypotheses do not differ by a large amount in any bin,
\begin{align}
\label{eq:hypothesis-test-small-difference-condition}
    |\delta^i|
    \ll
    1
    \qquad\text{for all $i$, where}\qquad
    \delta^i
    &=
    \frac{H_\text{SM}^i - H_\text{BSM}^i}{H_\text{BSM}^i}
    \,.
\end{align}
The test statistic for the Asimov data set, $A = H_\text{SM} = H(0)$, is then
\begin{align}
    q_\text{A}
    &=
    \max_{0 \leq \mu \leq 1}
    \left(
    -2 \log\left(
    \frac
    {P(A|H_\text{BSM})}
    {P(A|H(\mu))}
    \right)
    \right)
    \\
    &=
    -2 \log\left(
    \frac
    {P(A|H_\text{BSM})}
    {P(A|H(0))}
    \right)
    \\
    &=
    -2 \sum_{i=1}^2
    \log
    \left(
    \left(
    \frac{H_\text{BSM}^i}{H(0)^i}\right)^{A^i}
    \frac{e^{-H_\text{BSM}^i}}{e^{-H(0)^i}}
    \right)
    \\
    &=
    -2 \sum_{i=1}^2
    \left(
    A^i\log
    H_\text{BSM}^i
    - A^i \log H(0)^i
    - H_\text{BSM}^i + H(0)^i
    \right)
    \\
    &=
    -2 \sum_{i=1}^2
    \bigg(
    H_\text{BSM}^i (1 + \delta^i)\log H_\text{BSM}^i
    - H_\text{BSM}^i (1 + \delta^i)\log\left( H_\text{BSM}^i (1 + \delta^i) \right) + 
    \\
    &\hspace{2cm}
    - H_\text{BSM}^i + H_\text{BSM}^i (1 + \delta^i)
    \bigg)
    \nonumber \\
    &=
    -2 \sum_{i=1}^2
    \left(
    - H_\text{BSM}^i (1 + \delta^i)\log\left(1 + \delta^i \right)    
    + H_\text{BSM}^i \delta^i
    \right)
    \\
    &\approx
    -2 \sum_{i=1}^2
    \left(
    - H_\text{BSM}^i (1 + \delta^i)\left(\delta^i - \frac{(\delta^i)^2}{2}\right)    
    + H_\text{BSM}^i \delta^i
    \right)
    \\
    &\approx
    -2 \sum_{i=1}^2
    \left(
    - H_\text{BSM}^i \left(\delta^i + \frac{(\delta^i)^2}{2}\right)    
    + H_\text{BSM}^i \delta^i
    \right)
    \\
     &=
    -2 \sum_{i=1}^2
    \left(
    - H_\text{BSM}^i \frac{(\delta^i)^2}{2}
    \right)
    \\
    &=
    \sum_{i=1}^2
    \frac{(A^i - H_\text{BSM}^i)^2}{H_\text{BSM}^i}
    \,,
\end{align}
where we have used the fact that the test statistic is maximised by $\mu=0$ for the Asimov data set (since $H(0) = H_\text{SM} = A$) and $\log(1+\delta) \approx \delta - \delta^2/2$.  As in the goodness-of-fit test, the test statistic of the Asimov data set can again be approximated by \textbf{Pearson's $\chi^2$ test statistic}
\begin{align}
    q_\text{A} \approx \chi^2_\text{A}
    &=
    \sum_{i=1}^2
    \frac{(A^i - H_\text{BSM}^i)^2}{H_\text{BSM}^i}
    \\
    &=
    3.98
    \,.
\end{align}
We can then estimate the $p$-value for the hypothesis test using
\begin{align}
    p
    &=
    \int_{\chi^2_\text{A}}^\infty \frac{1}{2}f(q;1)\,dq
    \\
    &=
    \int_{\chi^2_\text{A}}^\infty \frac{1}{2}
    \frac{e^{-\frac{q}{2}}}{\sqrt{2\pi q}}
    \,dq
    \\
    &=
    1-
    \Phi\left(\sqrt{\chi^2_\text{A}}\right)
    \\
    &=
    \frac{1}{2}
   \text{erfc}\left(\sqrt{\frac{\chi^2_\text{A}}{2}}\right)
    \\    
    &\approx
    0.023
    \,,
\end{align}
where $\text{erfc}$ is the complementary error function. Again, this is in good agreement with the exact $p$-value in \cref{eq:exact-p-value-poisson-hypothesis-test}.

The different $p$-values for the goodness-of-fit test and the hypothesis test come from the different distributions of test statistics ($f(q;2)$ for the goodness-of-fit test and $f_\frac{1}{2}(q;1)$ for the hypothesis test) rather than from the test statistic of the Asimov data set, which in both cases is approximately Pearson's $\chi^2$ test statistic ($\chi^2_\text{A}$).  Note that in the goodness-of-fit test the $k=2$ in $f(q;2)$ was given by the number of bins (which is specific to our particular 2-bin example), while in the hypothesis test the $k=1$ in $f_\frac{1}{2}(q;1)$ was related to the one parameter family of hypotheses, $H(\mu)$ (which can always be assumed independently of the specific binning).

\subsection{Realistic Example: Heavy Resonances at a Muon Collider}
\label{sec:realistic-example}

As a realistic example of Poisson-distributed data, we consider a heavy resonance search at a future muon collider. A muon collider is an attractive proposal because it combines a high centre-of-mass energy with a comparatively clean collision environment \cite{
Delahaye:2019omf,
mu_smash,
toward_muc,
muon_accelerator_program,
Black:2022cth,
MuonCollider:2022xlm,
Narain:2022qud,
P5:2023wyd,
InternationalMuonCollider:2024jyv,
InternationalMuonCollider:2025sys,
Begel:2025ldu}. 
One of its technical challenges is the short muon lifetime and current R\&D aims to address this by approximately 2040 \cite{InternationalMuonCollider:2024jyv}. Consequently, no experimental data are available at present. It is nevertheless important to produce sensitivity projections now, both to quantify the collider’s anticipated discovery reach and to enable meaningful comparisons with other future collider proposals. 

Heavy vector resonances are well-motivated BSM candidates, arising both in weakly coupled gauge extensions of the Standard Model and in strongly coupled composite-Higgs scenarios. For concreteness, we consider the Standard Model plus a colour-neutral $su(2)_L$ vector triplet with zero hypercharge, $V$, originating from a weakly coupled gauge extension to the SM gauge group (called model A in  \cite{Pappadopulo:2014qza}). The sensitivity of a muon collider to this model has been studied in Ref.~\cite{Acanfora:2026aay}.

\begin{figure}
\centering
\includegraphics[width=0.5\linewidth]{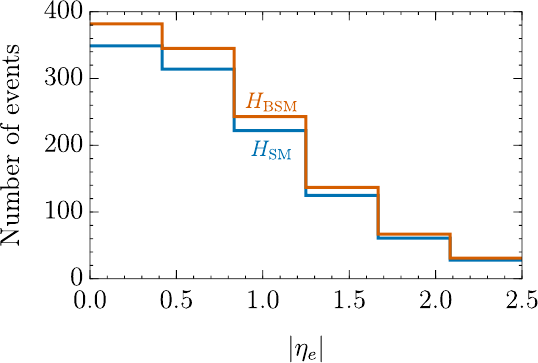}
\caption{Differential cross-section as a function of the absolute pseudorapidity of the final state electrons, $|\eta_e|$. The blue line shows the SM prediction and the orange line shows the heavy vector triplet model prediction for $m_V = 1.9$\,TeV and $g_V = 1.9$.}
\label{fig:example-P-observation}
\end{figure}

In this example, we consider the process $\mu^+\mu^- \to V^0 \to e^+ e^-$ at a $10\,$TeV muon collider delivering a total luminosity of $1\,$ab$^{-1}$. We compute the tree-level cross-section analytically, which depends on the resonance mass, $m_V$, and coupling strength, $g_V$. We include interference with the SM amplitude, and use the pseudorapidity distribution to discriminate between the SM and BSM hypotheses. The resulting spectra in pseudorapidity, $\eta$, for the SM and for a benchmark point $(m_V, g_V)=(1.9\,\text{TeV},\,1.9)$ are
\begin{align}
H_\text{SM} 
&=
(349,314,222,125,61.4,28.2) \\
H_\text{BSM} 
&=
(382,345,243,137,67.3,30.9)
\,.
\end{align}
with a uniform binning for $0\leq\abs{\eta}_e \leq 2.5$, shown in \cref{fig:example-P-observation}. The expected number of events at the muon collider is obtained from analytic cross sections multiplied by the luminosity and efficiency factors, where for these final state leptons we assume efficiency 1. Note that when choosing the binning it should not be taken to be finer than the experimental resolution.\footnote{Ref.~\cite{Acanfora:2026aay} also demonstrates that increasing the number of bins can increase the sensitivity of the analysis, but that there are diminishing returns and the important thing is that the shape differences between the two hypotheses are captured.} 

To obtain sensitivity projections, we assume the Asimov data set, i.e., we assume the observation coincides with the Standard Model expectation,
\begin{align}
    A = H_\text{SM} 
    \,.
\end{align}

We now compare a goodness-of-fit test and a hypothesis test. In this example there are so many events and bins that exact computation of the $p$-values is impractical, so  we estimate them using the Monte Carlo and asymptotic procedures described in \cref{subsec:P-MC-goodness,subsec:P-MC-hypothesis-test}.

\begin{figure}
\centering
\includegraphics[width=0.49\textwidth]{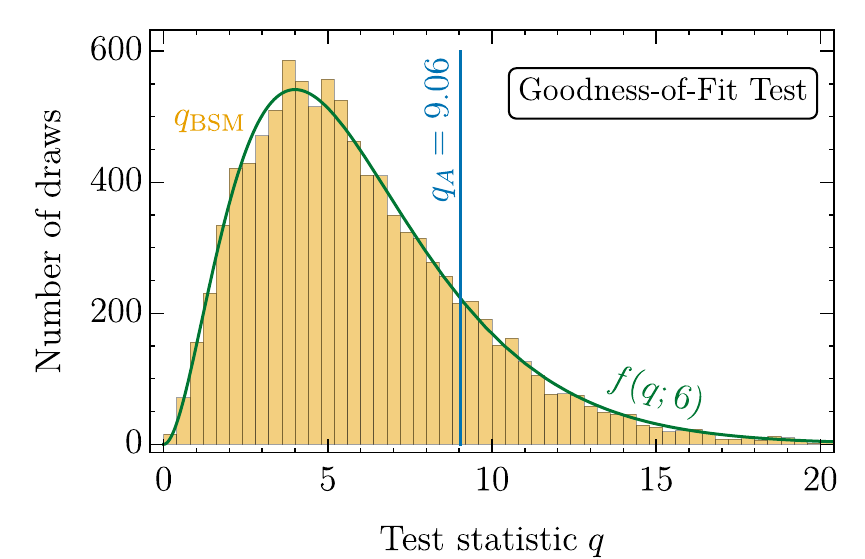}
\includegraphics[width=0.49\textwidth]{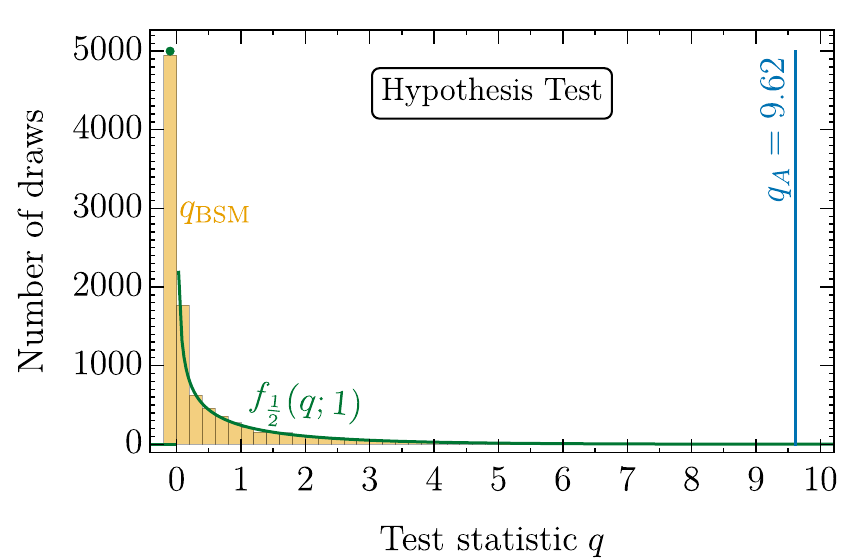}
\caption{The distribution of the test statistic obtained by Monte-Carlo sampling $H_\text{BSM}$ using a goodness-of-fit test (left) and a hypothesis test (right). The green lines show the asymptotic limits $f(q;6)$ (left) and $f_{\frac{1}{2}}(q;1)$ (right). The blue vertical lines show the test statistic of the Asimov data set, $q_\text{A}$.}
\label{fig:example-P-test-statistic}
\end{figure}

\Cref{fig:example-P-test-statistic} (left) shows the distribution of the goodness-of-fit test statistic, defined in \cref{eq:test-statistic-P-goodness-of-fit}, obtained from $10^4$ draws from $H_{\text{BSM}}$ (assuming that the data will be Poisson distributed). The green curve shows the corresponding asymptotic description $f(q;6)$ from \cref{eq:chi2-distribution}, i.e., a $\chi^2$ distribution with $k=6$ degrees of freedom. The blue vertical line shows the value of the test statistic for the Asimov data set, $q_A = 9.06$.  We find a $p$-value of $p = 1613/10^4 = 0.16$.  Using the asymptotic approximation for the distribution of the test statistic we find
\begin{align}
    p 
    &=
    \int_{q_A}^\infty f(q;6) \,dq
    \\&=
    0.17
\end{align}
and using the Pearson's $\chi^2$ test statistic of $\chi_A^2 = 9.33$ gives
\begin{align}
    p 
    &=
    \int_{\chi_A^2}^\infty f(q;6) \,dq
    \\&=
    0.16
    \,.
\end{align}
We see good agreement between these approaches, which indicate that the experiment could not rule out $H_\text{BSM}$ at the 95\% CL.

In \cref{fig:example-P-test-statistic} (right) we show the distribution of the hypothesis-test test statistic using a maximum likelihood estimator defined in \cref{eq:test-statistic-P-hypothesis}, obtained from $10^4$ Monte Carlo draws from $H_{\text{BSM}}$ assuming Poisson distributed data. We overlay the corresponding asymptotic approximation $f_\frac{1}{2}(q;1)$ discussed in \cref{eq:half-chi-squared} in green, i.e., a $\tfrac{1}{2}\chi^2$ distribution with one degree of freedom, and the test statistic for the Asimov data set, $q_A = 9.62$.  Since the $p$-value is given by the fraction of Monte Carlo draws with a test statistic $q \ge q_A$, we see that the goodness-of-fit distribution (left) contains a much larger tail beyond $q_A$ than the hypothesis-test distribution (right), leading to a larger $p$-value.  Using Monte Carlo methods, for the hypothesis test we find $p = 13/10^4 = 0.0013$.  Using the asymptotic approximation for the distribution of the test statistic we find
\begin{align}
    p 
    &=
    \int_{q_A}^\infty f_\frac{1}{2}(q;1) \,dq
    \\&=
    0.00096
\end{align}
and using the Pearson's $\chi^2$ test statistic of $\chi_A^2 = 9.33$ gives
\begin{align}
    p 
    &=
    \int_{\chi_A^2}^\infty f_\frac{1}{2}(q;1) \,dq
    \\&=
    0.0011
    \,.
\end{align}
We see good agreement between these approaches, which indicate that the experiment \textit{could} rule out $H_\text{BSM}$ at the 95\% CL.  We clearly see that the two different approaches give very different conclusions, and this is because in the two cases very different statistical questions are being asked.  The goodness-of-fit test simply asks how likely is it to get $A$ by chance when $H_\text{BSM}$ is the true model of nature, while the hypothesis test (roughly) asks whether it is more likely for $A$ to come from $H_\text{SM}$ or from $H_\text{BSM}$.

\begin{figure}
\centering
\includegraphics[width=0.55\textwidth]{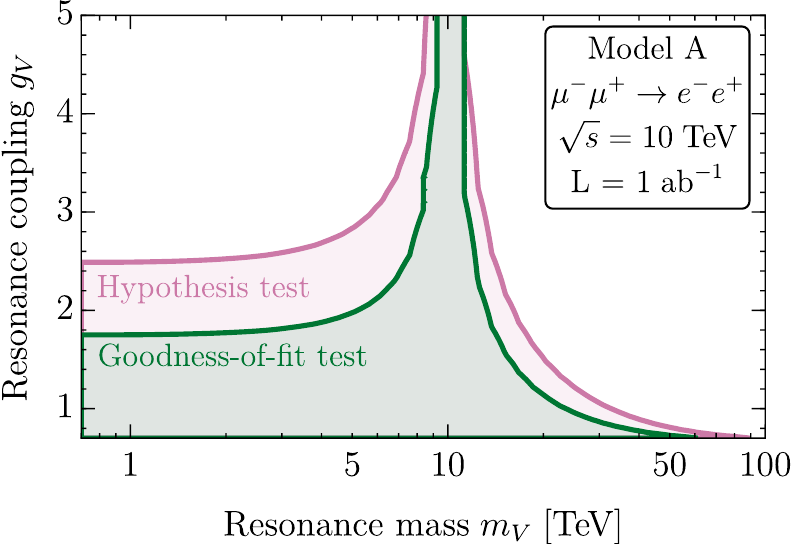}
\caption{Projected 95\% CL sensitivity of a goodness-of-fit test (green) and a hypothesis test (pink).}
\label{fig:P-comparison}
\end{figure}

\Cref{fig:P-comparison} shows the projected 95\% CL sensitivity of a goodness-of-fit test (green) and a hypothesis test (pink) in the BSM parameter space $(m_V, g_V)$. Each curve is obtained by evaluating the test statistic of the Asimov data set (\cref{eq:p-qa-asimov-gof} and \cref{eq:qa-asimov-ht-wilks}) across a grid of BSM parameter points and requiring $q_A > 12.6$ for goodness-of-fit and $q_A > 2.71$ for the hypothesis test.\footnote{Note that requiring $q_A > 3.84$ for the hypothesis test and including systematic errors provides 95\% CLs exclusion limits which agree with the results obtained in fig.~3.9 in Ref.~\cite{Acanfora:2026aay}.} This corresponds to $p<0.05$ for a $\chi^2$ distribution with six degrees of freedom (for the goodness-of-fit test) and $p<0.05$ for a half $\chi^2$ distribution with one degree of freedom (for the hypothesis test). We verified that using $\chi^2_A$ instead of $q_A$ for either of the statistical tests leads to essentially the same sensitivity projections.

We see that the contours in \cref{fig:P-comparison} are very different, with the hypothesis test excluding a larger portion of parameter space.  Since the experimental community typically uses hypothesis tests, this is what should be used when making projections. Comparing goodness-of-fit tests with hypothesis tests is not a like-for-like comparison.  

We have also checked that including uncorrelated systematic errors using \cref{eq:es-chi-squared} and using the method outlined in Refs.~\cite{Conway:2011in,Acanfora:2026aay} give almost identical contours in \cref{fig:P-comparison}.

\section{Multinomially Distributed Data}
\label{sec:multinomial}

We now compute exact $p$-values for both a goodness-of-fit test and a hypothesis test for multinomially distributed data. While the main ideas are a repeat of those discussed in \cref{sec:Poisson}, we feel it is invaluable for researchers to see examples of how to extend the approach to other probability distributions.  Data follows a multinomial distribution when bins are statistically independent (i.e., the errors are statistically dominated) but the total event count is fixed.

A concrete example of multinomially distributed data is the observation of an exploding black hole~\cite{Baker:2021btk,Baker:2025ffi}, where the distance to the explosion is not known. While BSM physics can modify both the spectral shape and the total number of detected photons, the unknown distance prevents us from using the absolute number of photons directly. Instead, we normalize the photon distribution, assuming a future observation with $N_\gamma$ photons, and use the shape information alone to discriminate between the SM and BSM hypotheses. We will come back to this example in \cref{sec:realistic-example-m}.

\begin{figure}
\centering
\includegraphics[width=0.5\linewidth]{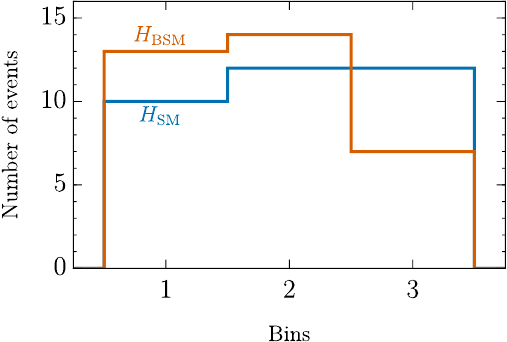}
\caption{The SM and BSM predictions, $H_\text{SM}$ (blue) and $H_\text{BSM}$ (orange), for our three bin multinomially distributed example.}
\label{fig:2-bin-multinomial-data}
\end{figure}

Here we consider a toy example and assume that a total number of $N = 34$ events is distributed over three bins.\footnote{Note that while the total number of events can in principle be different for $H_\text{SM}$ and $H_\text{BSM}$, we here assume both are the same for simplicity.} We imagine that the Standard Model predicts the following bin counts
\begin{align}
    H_\text{SM} = (10,12,12)
    \,,
\end{align}
while the Standard Model plus some BSM physics, at a particular point in the BSM parameter space, predicts
\begin{align}
    H_\text{BSM} = (13,14,7)
    \,.
\end{align}
We plot these distributions in \cref{fig:2-bin-multinomial-data}.  Note that with multinomially distributed data, three bins have two degrees of freedom, since the third bin is determined by the constraint $o_3 = N - o_1 - o_2$.

\subsection{Exact $p$-value}
\label{sec:exact-p-value-m}

In this section we will compute the exact $p$-value for this three-bin toy example for a goodness-of-fit and a hypothesis test.

The probability of observing $o_1$, $o_2$ and $o_3$ events in the three bins, with the condition $\sum_{i=1}^3 o_i = N$, if $H_\text{BSM}$ is the true model of nature, is
\begin{align} \label{eq:multilikelihood}
    P_M
    &=
    \frac{N!}{o_1!o_2!o_3!}
    p_1^{o_1} p_2^{o_2} p_3^{o_3}
    \,,
\end{align}
where
\begin{align}
    p_i &= \frac{H_\text{BSM}^i}{N} 
\end{align}
is the probability of an event falling in the $i$-th bin. The general probability mass function for a BSM hypothesis which predicts $H_\text{BSM}$ events in $n_B$ bins and an observation consisting of $o_i$ events is
\begin{align} \label{eq:multi-pmf}
    P_M (o|H_\text{BSM})
    &=
    N! \prod_{i=1}^{n_B} \frac{p_i^{o_i}}{o_i!}
    \,, \qquad \text{with} \qquad p_i = \frac{H_\text{BSM}^i}{N} \,.
\end{align}
This is called the \textbf{likelihood} and is the multinomial analogue of \cref{eq:poisson-pmf}.

As before we can now ask how likely it is that we would observe events consistent with $H_\text{SM}$, i.e., the Asimov data set, if $H_\text{BSM}$ were the true model of nature.

\subsubsection{Goodness of Fit}
\label{sec:M-exact-goodness}

The probability of observing the Asimov data set (which is consistent with the SM hypothesis, $A = H_\text{SM}$) if $H_\text{BSM}$ were the true model of nature is
\begin{align}
    P_M^A
    &=
    \frac{N!}{A_1!A_2!A_3!}
    p_1^{A_1} p_2^{A_2} p_3^{A_3}
    \\&=
    \frac{34!}{10!\, 12! \,12!}
    \left(\frac{13}{34}\right)^{10} 
    \left(\frac{14}{34}\right)^{12} 
    \left(\frac{7}{34}\right)^{12} 
    \\&\approx
    0.0033
    \,,
\end{align}
while the probability of observing events equal to the BSM hypothesis is
\begin{align}
    P_M^\text{BSM}
    &=
    \frac{N!}{H_\text{BSM}^1!H_\text{BSM}^2!H_\text{BSM}^3!}
    p_1^{H_\text{BSM}^1} p_2^{H_\text{BSM}^2} p_3^{H_\text{BSM}^3}
    \\
    &=
    \frac{34!}{13!\, 14!\, 7!}
    \left(\frac{13}{34}\right)^{13} 
    \left(\frac{14}{34}\right)^{14} 
    \left(\frac{7}{34}\right)^{7} 
    \\&\approx
    0.025
    \,.
\end{align}
This is the most likely outcome of the experiment and the probability of all other outcomes must be less than 0.025.

\begin{figure}
\centering
\includegraphics[width=0.49\linewidth]{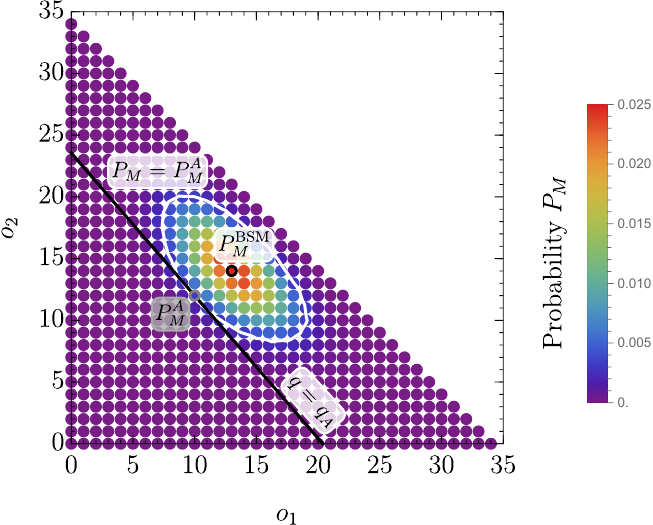}
\includegraphics[width=0.49\linewidth]{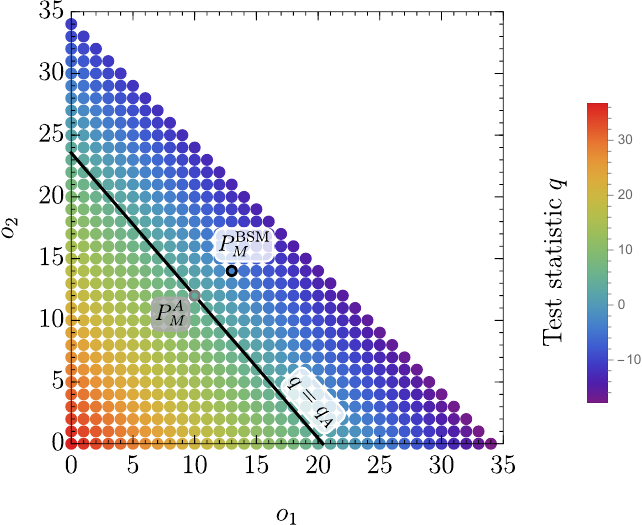}
\caption{Left: The probabilities for different outcomes in the two bins given $H_\text{BSM}$ is the true model of nature for multinomially distributed data, with the contours $P_M = P_M^\text{A}$ (white) and $q = q_\text{A}$ (black).  Right: The corresponding test statistics.}
\label{fig:exactPValueMultinomialPlot}
\end{figure}

In \cref{fig:exactPValueMultinomialPlot} (left) we show the probabilities of all possible ways of distributing $N = 34$ events among three bins, assuming $H_\text{BSM}$ is the true model of nature. Each outcome is labeled by the counts in the first two bins, $(o_1,o_2)$, with the third bin fixed by $o_3=N-o_1-o_2$. Unlike the Poisson case in \cref{fig:exactPValuePoissonGoodnessofFitPlot}, only the lower half of the $(o_1,o_2)$ plane is physically allowed because the total event count is fixed. Outcomes near the $H_{\text{BSM}}$ expectation (black circle) have the largest probability, while more distant outcomes are increasingly unlikely. The small gray circle marks the outcome corresponding to the Asimov data set and the white contour shows where the probabilities are equal to $P_M^\text{A} \approx 0.0033$. 
As before, the $p$-value for the goodness-of-fit test is obtained by summing the probabilities of all outcomes on or outside the white contour, i.e., all points with $P_M(o_1,o_2)\le P_M^{\text{A}}$. The $p$-value is
\begin{align}
\label{eq:exact-p-value-multinomial-goodness-of-fit}
    p &=
    \sum_{\substack{o_1,o_2 \text{ with } \\P_M(o_1,o_2) \leq P_M^\text{A}}} P_M(o_1,o_2)
    \\
    &=
    0.12
    \,.
\end{align}
This $p$-value tells us that the observation is not statistically significant at the 5\% level, i.e., $H_\text{SM}$ could have been observed by chance alone if $H_\text{BSM}$ is the true model of nature.  The corresponding confidence level is CL $ = 1-p = 0.88$ and a future observation consistent with $H_\text{SM}$ would not exclude $H_\text{BSM}$ at the $95\%$ confidence level.

\subsubsection{Hypothesis Test}

In the goodness-of-fit test we implicitly used the probability of each outcome assuming $H_\text{BSM}$ as the test statistic.  For a hypothesis test, the test statistic is instead the log-likelihood ratio of the probabilities assuming the two competing hypotheses (where we assume an equal number of total events for each hypothesis),
\begin{align}
\label{eq:multi-hypo-test-statistic-1}
    q
    &=
    -2 \log\left(
    \frac
    {P(o|H_\text{BSM})}
    {P(o|H_\text{SM})}
    \right)
    \\
    &=
    -2 \log\left(
    \left(\frac{N!}{N!}\right)
    \prod{i =1}^3
    \frac{\left(\frac{H_\text{BSM}^i}{N}\right)^{o_i}}{\left(\frac{H_\text{SM}^i}{N}\right)^{o_i}}
    \left(
    \frac{o_i!}{o_i!}
    \right)
    \right)
    \\
    &= \label{eq:multi-hypo-test-statistic}
    -2 
    \sum_{i=1}^3
    o_i
    \log\left(
    \frac{H_\text{BSM}^i}{H_\text{SM}^i}
    \right)
    \,.
\end{align} 
We show the test statistic across the possible outcome space $(o_1,o_2)$ in \cref{fig:exactPValueMultinomialPlot} (right). For the Asimov data set (the grey circle), the test statistic is $q_{\text{A}}=3.89$, and the black contour in the left panel shows where $q(o_1,o_2)=3.89$.
To obtain the hypothesis-test $p$-value, we sum the probabilities of all outcomes that have a test statistic more extreme than that of the Asimov data set, i.e., those with $q(o_1,o_2)\ge q_{\text{A}}$. Geometrically, these correspond to the points lying to the lower left of the black contour,
\begin{align}
\label{eq:exact-p-value-multinomial-hypothesis-test}
    p &=
    \sum_{\substack{o_1,o_2 \text{ with } \\ q(o_1,o_2) \ge q_\text{A}}} P(o_1,o_2)
    \\
    &=
    0.024
    \,.
\end{align}
This corresponds to a confidence level of CL $=0.976$. Thus, as in \cref{sec:exact-p-value-m}, we find that while the goodness-of-fit test does not exclude the BSM point at 95\% confidence, the hypothesis test does.  This is because the two tests answer fundamentally different questions, and we advocate for the use of hypothesis tests in BSM phenomenology.

\subsection{Monte Carlo and Asymptotic Estimates of the $p$-value}
\label{sec:approximate-p-value-m}

In many realistic applications, the number of events and bins, and hence the number of possible outcomes, is too large to compute exact $p$-values by explicit enumeration. We can then turn to Monte Carlo and asymptotic approximations. For concreteness, we continue to illustrate these methods using the three-bin toy example introduced above, with $N=34$, 
\begin{align}
    H_\text{SM} = (10,12,12)
    \,,
\label{eq:exMultiDistSM}
\end{align}
and
\begin{align}
    H_\text{BSM} = (13,14,7)
    \,.
\label{eq:exMultiDistBSM}
\end{align}

\subsubsection{Goodness of Fit}
\label{subsec:M-MC-gof}

We first consider the goodness-of-fit test, and approximate the $p$-value with Monte Carlo methods before looking at some asymptotic approximations.

\paragraph{Monte Carlo Methods:}
%
\begin{figure}
\centering
\includegraphics[width=0.49\linewidth]{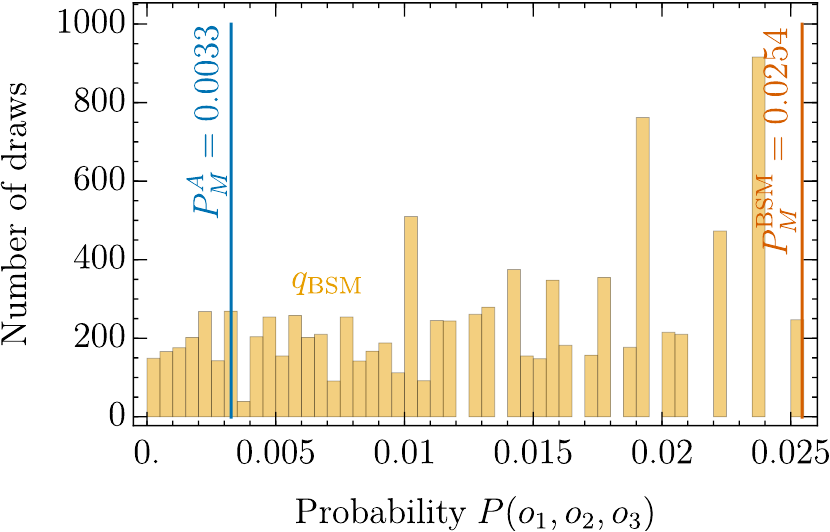}
\includegraphics[width=0.49\linewidth]{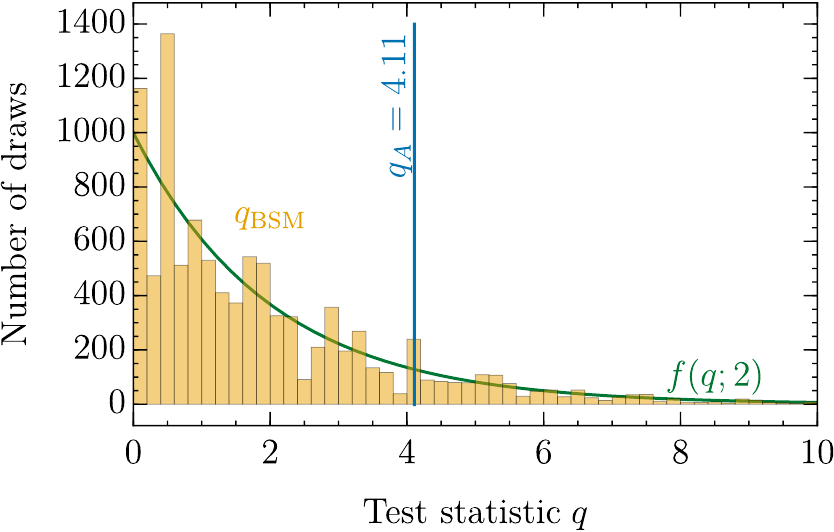}
\caption{Histograms of the probabilities (left) and the test statistic (right) obtained from $10^4$ draws from $H_\text{BSM}$, along with the probability and test statistic for the Asimov data set $A = H_\text{SM}$ in blue.  The orange line shows the probability of observing $H_\text{BSM}$ and the green line shows a $\chi^2$ distribution with 2 degrees of freedom.}
\label{fig:MCMultiProbsPlot}
\end{figure}
%
As for Poisson distributed data, we take the probability under the null hypothesis ($H_\text{BSM}$) as the test statistic. We draw from the BSM hypothesis $N_{\text{MC}}=10^4$ times. Each trial consists of drawing an outcome $(o_1,o_2,o_3)$ from the multinomial distribution with probability $P(o_1,o_2,o_3)$, given in \cref{eq:multilikelihood}. The resulting distribution of $P(o_1,o_2,o_3)$ values for the Monte Carlo draws is shown in \cref{fig:MCMultiProbsPlot} (left). Recall the most probable outcome, $P_M^\text{BSM}$, has probability $0.025$, while $o = A = H_\text{SM}$ has probability $P_M^\text{A} = 0.0033$. As in \cref{subsec:P-MC-goodness}, we see that the more probable outcomes are drawn more often, but there are fewer of them.
The $p$-value of the goodness-of-fit test can be obtained by counting the number of draws with $P(o_1,o_2,o_3) \leq P_M^\text{A}$ and dividing by the number of Monte Carlo trials. The result for this particular set of draws is 
\begin{align}
    p &= 
    \frac{1208}{10^4}
    \\
    &\approx
    0.12
    \,,
\end{align}
which agrees with the exact method in \cref{sec:M-exact-goodness}.

As in the Poisson case, the detailed form of the probability distribution, \cref{fig:MCMultiProbsPlot} (left), is specific to this particular example and is not easily predicted in general. When there are large event counts, we can address this by using the same test statistic as in the Poisson case, \cref{eq:test-statistic-P-goodness-of-fit}, 
\begin{align} \label{eq:test-statistic-M-goodness-of-fit}
    q
    &=
    -2 \log\left(
    \frac
    {P(o|H_\text{BSM})}
    {P(H_\text{BSM}|H_\text{BSM})}
    \right)
    \\
    &=
    -2 \log\left(
    \left(\frac{N!}{N!}\right)
    \prod{i =1}^3
    \frac{\left(\frac{H_\text{BSM}^i}{N}\right)^{o_i}}{\left(\frac{H_\text{BSM}^i}{N}\right)^{H_\text{BSM}^i}}
    \left(
    \frac{H_\text{BSM}^i!}{o_i!}
    \right)
    \right)
    \\
    &=
    -2 
    \sum_{i =1}^3 \left(
    (o_i - H_\text{BSM}^i) \log \left( \frac{H_\text{BSM}^i}{N}\right)
    + \log (H_\text{BSM}^i!) - \log (o_i!)
    \right)
    \,,
\end{align}
which gives the distribution shown in \cref{fig:MCMultiProbsPlot} (right).  The value of the test statistic of the Asimov data set $A = H_\text{SM}$ is 
\begin{align}
\label{eq:test-statistic-asimov-M-goodness-of-fit}
    q_\text{A} 
    &=
    -2 
    \sum_{i =1}^3 \left(
    (A^i - H_\text{BSM}^i) \log \left( \frac{H_\text{BSM}^i}{N}\right)
    + \log(H_\text{BSM}^i!) - \log(A^i!)
    \right)
    \\
    &= 
    4.11
    \,.
\end{align}
To estimate the $p$-value we count the number of draws more extreme than the Asimov data set (that is, to the right of the blue line in \cref{fig:MCMultiProbsPlot}, right panel) and divide by the number of draws, again giving
\begin{align}
    p &= 
    \frac{1208}{10^4}
    \\
    &\approx
    0.12
    \,.
\end{align}

\paragraph{Asymptotic Distribution of the Test Statistic:}
While the distribution in \cref{fig:MCMultiProbsPlot} (left) was difficult to predict, the distribution of the test statistics in  \cref{fig:MCMultiProbsPlot} (right) is well fitted by a $\chi^2$ distribution with $k=2$ degrees of freedom $f(q;2)$, as given in \cref{eq:chi2-distribution-k=2}.  Note that even though there are three bins in this example, there are only two degrees of freedom since the third bin is fixed by $o_3 = N - o_1 - o_2$.  In general, as the number of events increases, the distribution asymptotes to a $\chi^2$ distribution with $k=$ (number of bins $-1$) degrees of freedom.  We can therefore estimate the $p$-value by integrating $f(q;2)$ from $q_{\rm A}$ to $\infty$,
\begin{align} \label{eq:p-value-M-goodness-of-fit-asymptotic}
    p
    &\approx
    \int_{q_\text{A}}^\infty f(q;2)\,dq
    \\
    &=
    e^{-\frac{q_\text{A}}{2}}
    \\
    &\approx
    0.13
    \,.
\end{align}
We see that we can estimate the $p$-value without explicit enumeration of all outcomes (to compute the $p$-value exactly) and without performing a Monte Carlo simulation.  As the total number of events get larger, this is expected to give a better approximation to the exact $p$-value.

\paragraph{Asymptotic Limit of the Test Statistic of the Asimov Data Set:}
We now consider the limits where the two hypotheses do not differ by a large amount in any bin,
\begin{align}
\label{eq:largelimit1}
    |\delta^i|
    \ll
    1
    \qquad\text{for all $i$, where}\qquad
    \delta^i
    &=
    \frac{A^i-H_\text{BSM}^i}{H_\text{BSM}^i}
    \,,
\end{align}
and where there is a large number of events in each bin,
\begin{align}
\label{eq:largelimit2}
    A^i
    \gtrsim
    10
    \,.
\end{align}
The multinomial constraint of a fixed total event count implies that
\begin{align} \label{eq:multi-constraint}
    \sum_i \left(A^i-H_\text{BSM}^i \right) =\sum_i H_\text{BSM}^i \delta^i = 0 \,.
\end{align}

Using Stirling's approximation $\log(n!) \approx n\log n - n$ and $\log(1+\delta) \approx \delta - \delta^2/2$, we can derive an approximate expression for the test statistic of the Asimov data set,
\begin{align}
    q_\text{A}
    &=
     -2 
    \sum_{i =1}^3 \left[
    (A^i - H_\text{BSM}^i) \log \left( \frac{H_\text{BSM}^i}{N}\right)
    + \log(H_\text{BSM}^i!) - \log(A^i!)
    \right]
    \\
    &\approx
    -2 \sum_{i=1}^3
    \bigg[
    A^i
    \log
    \left(
    H_\text{BSM}^i
    \right)
    -
    H_\text{BSM}^i
    \log
    \left(
    H_\text{BSM}^i
    \right)
    - A^i \log N + H_\text{BSM}^i \log N
    \\
    &\hspace{2cm}
    +
    H_\text{BSM}^i
    \log
    \left(H_\text{BSM}^i
    \right)
    -
    H_\text{BSM}^i
    -
    A^i
    \log
    \left(A^i
    \right)
    +
    A^i
     \nonumber\bigg]
    \\
     &=
    -2 \sum_{i=1}^3
    \bigg[
    H_\text{BSM}^i (1 + \delta^i)
    \log
    \left(
    H_\text{BSM}^i
    \right)
    -
    H_\text{BSM}^i
    \log
    \left(
    H_\text{BSM}^i
    \right)
     \\
    &\hspace{2cm}
    - H_\text{BSM}^i (1 + \delta^i) \log N + H_\text{BSM}^i \log N
    +
    H_\text{BSM}^i
    \log
    \left(H_\text{BSM}^i
    \right)
    -
    H_\text{BSM}^i
    \nonumber \\
    &\hspace{2cm}
    - H_\text{BSM}^i (1 + \delta^i)
    \log
    \left(H_\text{BSM}^i (1 + \delta^i)
    \right)
    +
    H_\text{BSM}^i (1 + \delta^i)
    \bigg]
    \nonumber \\
    &=
    -2 \sum_{i=1}^3
    \bigg[
    H_\text{BSM}^i \delta^i
    \log
    \left(
    H_\text{BSM}^i
    \right)
    +
    H_\text{BSM}^i
    \log
    \left(H_\text{BSM}^i
    \right)
    \\
    &\hspace{2cm}   
    -
    H_\text{BSM}^i (1 + \delta^i)
    \log
    \left(H_\text{BSM}^i
    \right)
    -
    H_\text{BSM}^i (1 + \delta^i)
    \log
    \left(1 + \delta^i
    \right)
    \bigg]
     \nonumber \\
    &=
    -2 \sum_{i=1}^3
    \bigg[
    -
    H_\text{BSM}^i (1 + \delta^i)
    \left(\delta^i - \frac{\left(\delta^i \right)^2}{2}
    \right)
    \bigg]
    \\
    &\approx
    -2 \sum_{i=1}^3
    \bigg[
    -
    H_\text{BSM}^i
    \left(\delta^i + \frac{\left(\delta^i \right)^2}{2}
    \right)
    \bigg]
    \\
     &=
    -2 \sum_{i=1}^3
    \bigg[
    -
    H_\text{BSM}^i
    \frac{\left(\delta^i \right)^2}{2}
    \bigg]
    \\
    &=
    \sum_{i=1}^3
    \frac{(A^i-H_\text{BSM}^i)^2}{H_\text{BSM}^i}
    \,,
\end{align}
where we have used the condition \cref{eq:multi-constraint} several times. This is again \textbf{Pearson's $\chi^2$ test statistic}.

In our three-bin toy example we find
\begin{align}
    q_\text{A} \approx \chi^2_\text{A}
    &=
    \sum_{i=1}^3
    \frac{(A^i-H_\text{BSM}^i)^2}{H_\text{BSM}^i}
    \\
    &=
    4.55
    \,.
\end{align}
Using the observation that the distribution of test statistics asymptotes to a $\chi^2$ distribution with 2 degrees of freedom, we can estimate the $p$-value using \textbf{Pearson's $\chi^2$ test},
\begin{align} \label{P-goodness-of-fit-Pearsons-chi2}
    p
    &=
    \int_{\chi^2_\text{A}}^\infty f(q;2)\,dq
    \\
    &=
    e^{-\frac{\chi^2_\text{A}}{2}}
    \\
    &\approx
    0.10
    \,.
\end{align}
Note that the Monte Carlo estimate of the $p$-value agrees well with the exact result obtained in the previous section, whereas Pearson’s $\chi^2$ approximation gives a noticeably smaller $p$-value. The discrepancy arises because the conditions required to derive the Pearson $\chi^2$ test statistic (i.e., sufficiently large expected counts in each bin and similar counts under both hypotheses) are only marginally satisfied for our toy multinomial examples (\cref{eq:exMultiDistSM} and \cref{eq:exMultiDistBSM}).

\paragraph{The $G$-test:}
Another test that is sometimes used for multinomially distributed data is the $G$-test.  Here, the test statistic (where traditionally $q$ is written as $G$) is
\begin{align} 
    G
    &=
    -2 \log\left(
    \frac
    {P(o|H_\text{BSM})}
    {P(o|o)}
    \right)
    \\
    &=
    -2 \log\left(
    \left(\frac{N!}{N!}\right)
    \prod{i =1}^3
    \frac{\left(\frac{H_\text{BSM}^i}{N}\right)^{o_i}}{\left(\frac{o_i}{N}\right)^{o_i}}
    \left(
    \frac{o_i!}{o_i!}
    \right)
    \right)
    \\
    &=
    -2 
    \sum_{i =1}^3 \left(
    o_i \log \left( \frac{H_\text{BSM}^i}{o_i}\right)
    \right)
    \,.
\end{align}
where the second $o$ in $P(o|o)$ is the {\bf maximum likelihood estimator} of some particular observation, $o$. That is, the hypothesis that best matches the observation, which is, by definition, $H=o$.  For the Asimov data set the test statistic is
\begin{align} 
\label{eq:test-statistic-M-goodness-of-fit-G-test}
    G_A
    &=
    -2 
    \sum_{i =1}^3 \left(
    A^i \log \left( \frac{H_\text{BSM}^i}{A^i}\right)
    \right)
    \\&=
    3.99
    \,.
\end{align}

\begin{figure}
\centering
\includegraphics[width=0.49\linewidth]{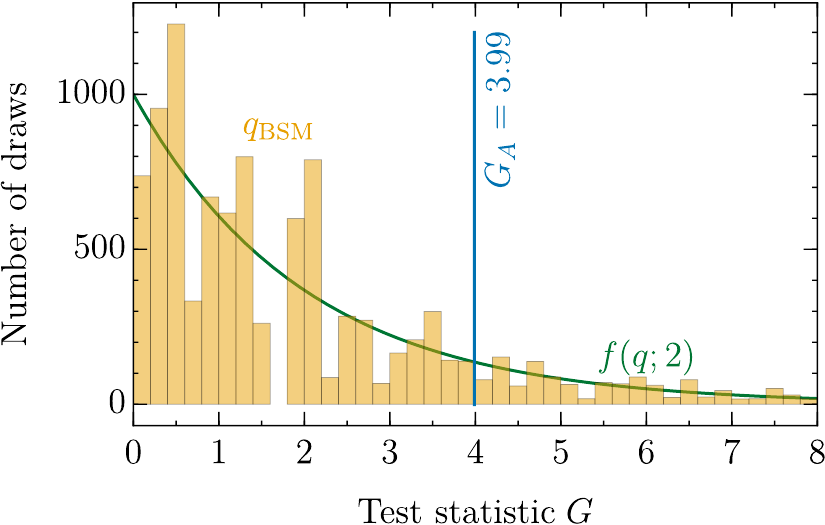}
\caption{Histogram of the test statistic obtained from $10^4$ draws from $H_\text{BSM}$ for the test statistic used in the $G$-test, along with the test statistic for the Asimov data set $A = H_\text{SM}$ (blue) and a $\chi^2$ distribution with 2 degrees of freedom (green).}
\label{fig:G-test}
\end{figure}

In \cref{fig:G-test} we plot the distribution of test statistics when drawing $N_\text{MC} = 10^4$ times from $H_\text{BSM}$.  As the number of events increases, this distribution also asymptotes to a $\chi^2$ distribution with $k=$ (number of bins $-1$) degrees of freedom (in this example, $k=2$).  Counting the number of MC events to the right of $G_A$ we find
\begin{align}
    p &= \frac{1428}{10^4}
    \\&= 0.14
    \,,
\end{align}
while using the asymptotic distribution we find
\begin{align}
    p &= \int_{G_A}^\infty f(q;2) \,dq
    \\&= 0.14
    \,.
\end{align}
We see that in both cases, for our example, this provides a good approximation to the exact $p$-value.  

Pearson's $\chi^2$ test discussed above can be derived as an approximation to the $G$-test under the same approximations as in \cref{eq:largelimit1} and \cref{eq:largelimit2} (i.e., the BSM hypothesis does not differ by a large amount from the Asimov data set in any bin and there is a large number of events in each bin). Using \cref{eq:largelimit1} and $\log(1+\delta) \approx \delta - \delta^2/2$, we can write
\begin{align}
    G_\text{A}
    &=
    -2 
    \sum_{i =1}^3 \left(
    A^i \log \left( \frac{H_\text{BSM}^i}{A^i}\right)
    \right)
    \\
    &=
    -2 
    \sum_{i =1}^3 \left(
    H_\text{BSM}^i (1 + \delta^i) \log \left( \frac{H_\text{BSM}^i}{H_\text{BSM}^i (1 + \delta^i)}\right)
    \right)
    \\
    &\approx
    -2 
    \sum_{i =1}^3 \left(
    - H_\text{BSM}^i (1 + \delta^i)  \left( \delta^i - \frac{(\delta^i)^2}{2}\right)
    \right)
    \\
    &\approx
    -2 
    \sum_{i =1}^3 \left(
    - H_\text{BSM}^i \frac{(\delta^i)^2}{2}
    \right)
    \\
    &=
    \sum_{i=1}^3
    \frac{(A^i-H_\text{BSM}^i)^2}{H_\text{BSM}^i}
    \\
    &=
    4.55
    \,,
\end{align}
where we used the condition \cref{eq:multi-constraint} in the penultimate step.  We see that the test statistic of the Asimov data set is the {\bf Pearson's $\chi^2$ test statistic}, and we can estimate the $p$-value using Pearson's $\chi^2$ test as described in \cref{P-goodness-of-fit-Pearsons-chi2} (since the asymptotic distributions of the test statistics are the same). Note that the $G$-test is often the preferred method for goodness-of-fit tests of multinomially distributed data as it is as simple to compute as Pearson's $\chi^2$ test but more accurate (as is the case for our example).

\subsubsection{Hypothesis Test}
\label{subsec:M-MC-hypothesis-test}

We now discuss Monte Carlo and asymptotic methods for hypothesis tests of multinomially distributed data.

\begin{figure}
\centering
\includegraphics[width=0.5\linewidth]{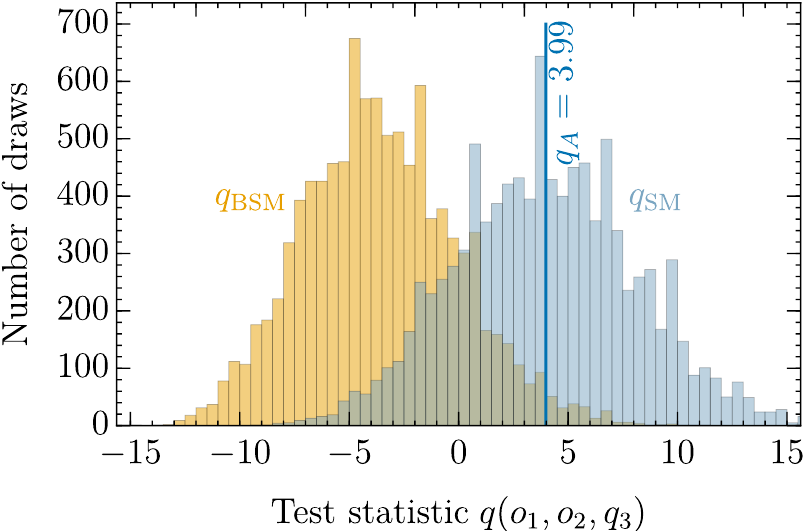}
\caption{Histograms of the test statistics obtained from $10^4$ draws from $H_\text{BSM}$ (orange) and from $H_\text{SM}$ (blue), along with the test statistic for the Asimov data set, $A =H_\text{SM}$.}
\label{fig:MCMultiTestStatisticPlot}
\end{figure}

\paragraph{Monte Carlo Methods:}
We now draw from both the SM and BSM hypotheses a large number of times, $N_\text{MC} = 10^4$, using probabilities given by the multinomial distribution, \cref{eq:multi-pmf}.  For each draw, $o$, we then compute the log-likelihood ratio test statistic given by \cref{eq:multi-hypo-test-statistic},
\begin{align}
    q
    &=
    -2 \log\left(
    \frac
    {P(o|H_\text{BSM})}
    {P(o|H_\text{SM})}
    \right)
    \\
    &=
    -2 
    \sum_{i=1}^3
    o_i
    \log\left(
    \frac{H_\text{BSM}^i}{H_\text{SM}^i}
    \right)
    \,.
\end{align}
The distribution of test statistics for the possible observations expected by chance from the SM and BSM hypotheses are shown in \cref{fig:MCMultiTestStatisticPlot}.  To find an estimate of the $p$-value we count the number of events drawn from $H_\text{BSM}$ with a test statistic more extreme than that of the Asimov data set,
\begin{align}
\label{eq:multi-hypothesis-test-test-statistic}
    q_\text{A} 
    &=
    -2 \sum_{i=1}^3\left(A^i\log
    \left(
    \frac{H_\text{BSM}^i}
    {A^i}
    \right)
    \right)
    \\
    &= 3.99
    \,,
\end{align}
giving
\begin{align}
    p &= 
    \frac{249}{10^4}
    \\
    &\approx
    0.025
    \,,
\end{align}
in agreement with the exact $p$-value computed in \cref{eq:exact-p-value-multinomial-hypothesis-test}.

\begin{figure}
\centering
\includegraphics[width=0.5\linewidth]{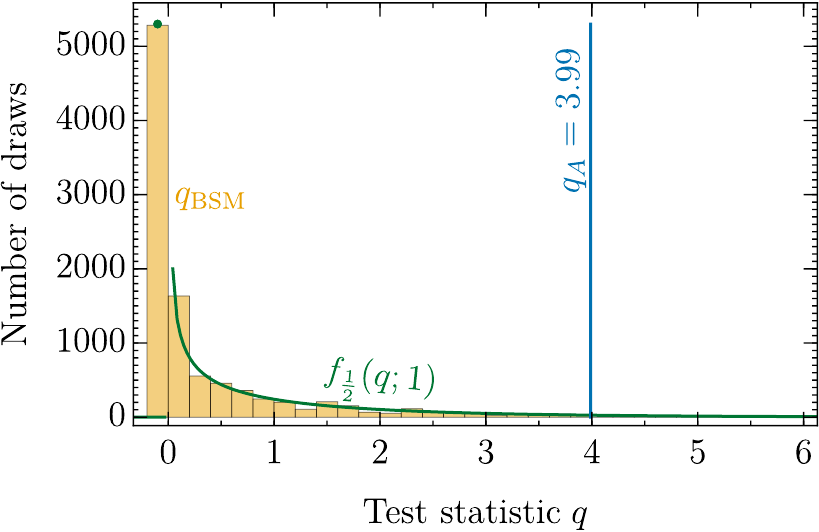}
\caption{Histogram of the test statistics obtained from $10^4$ draws from $H_\text{BSM}$ (orange) using the Wilks' theorem method, along with the test statistic for the Asimov data set, $A = H_\text{SM}$.  The green curve and dot illustrate a $\frac{1}{2} \chi^2$ distribution with 1 degree of freedom.}
\label{fig:MCMultiTestStatisticFCPlot}
\end{figure}

\paragraph{Wilks' Theorem:}
Unfortunately, the distributions shown in \cref{fig:MCMultiTestStatisticPlot} can not easily be approximated.  We now discuss a method for applying Wilks' theorem which does lead to an asymptotic approximation.  As in the Poisson case, we change the test statistic and replace the alternate hypothesis $H_\text{SM}$ with a one parameter family of hypotheses,
\begin{align}
    H(\mu) = \mu H_\text{BSM} + (1-\mu) H_\text{SM}
    \,,
\end{align}
which smoothly interpolate between $H_\text{SM}$ and $H_\text{BSM}$ as the parameter $\mu$ changes from 0 to 1.  The null hypothesis $H_\text{BSM}$ then lies in the space of possibilities of the alternate hypothesis $H(\mu)$.  We draw from $H_\text{BSM}$ a large number of times, $N_\text{MC} = 10^4$, and compute the test statistic,
\begin{align} \label{eq:test-statistic-M-hypothesis}
    q
    &=
    \max_{0 \leq \mu \leq 1}
    \left(
    -2 \log\left(
    \frac
    {P(o|H_\text{BSM})}
    {P(o|H(\mu))}
    \right)
    \right)
    \\
    &=
    \max_{0 \leq \mu \leq 1}
    \left(
    -2 \sum_{i=1}^3o_i\log
    \left(
    \frac{H_\text{BSM}^i}
    {H(\mu)^i}
    \right)
    \right)
    \,,
\end{align}
for each draw.  For each draw a different value of $\mu$ is found which maximises the test statistic, and $H(\mu)$ is then called the \textbf{maximum likelihood estimator}.  We show the distribution of test statistics in \cref{fig:MCMultiTestStatisticFCPlot}.

The test statistic for the Asimov data set is then
\begin{align}
    q_\text{A}
    &=
    \max_{0 \leq \mu \leq 1}
    \left(
    -2 \log\left(
    \frac
    {P(A|H_\text{BSM})}
    {P(A|H(\mu))}
    \right)
    \right)
    \\
\label{eq:multi-hypothesis-test-fc-test-statistic}
    &=
    -2 \sum_{i=1}^3\left(A^i\log
    \frac{H_\text{BSM}^i}
    {H(0)^i}
    \right)
    \\
    &=
    3.99
    \,,
\end{align}
since in this case $P(A|H(\mu))$ is maximised for $H(\mu) = A = H_\text{SM} = H(0)$.  This then agrees with the test statistic of the Asimov data set with the test statistic used above, \cref{eq:multi-hypothesis-test-test-statistic}.  We can then estimate the $p$-value by counting the number of events to the right of $q_\text{A}$ and dividing by the number of draws,
\begin{align}
    p &= 
    \frac{210}{10^4}
    \\
    &\approx
    0.021
    \,.
\end{align}
We see that this is in good agreement with the exact $p$-value.

\paragraph{Asymptotic Distribution of the Test Statistic:}
In \cref{fig:MCMultiTestStatisticFCPlot} we see that the distribution of test statistics can again be approximated by a $\frac{1}{2} \chi^2$ distribution with one degree of freedom, $f_{\frac{1}{2}}(q;1)$. We can then analytically estimate the $p$-value using,
\begin{align}
\label{eq:test-statistic-M-hypothesis-asymptotic}
    p
    &=
    \int_{q_\text{A}}^\infty f_{\frac{1}{2}}(q;1)\,dq
    \\
    &=
    \frac{1}{2}
   \text{erfc}\left(\sqrt{\frac{q_\text{A}}{2}}\right)
    \\
    &\approx
    0.023
    \,,
\end{align}
again in good agreement with the exact $p$-value.

\paragraph{Asymptotic Limit of the Test Statistic of the Asimov Data Set:}
For the Asmiov data set, we now consider the limits as in \cref{eq:largelimit1} and \cref{eq:largelimit2} where the two hypotheses do not differ by a large amount in any bin and there is a large number of events in each bin. Imposing the multinomial constraint of a fixed total event count implies that
\begin{align} 
\label{eq:multi-constraint2}
    \sum_i A^i - H_\text{BSM}^i = \sum_i H_\text{BSM}^i\delta^i = 0 \,.
\end{align}
Using $\log(1+\delta) \approx \delta - \delta^2/2$, we can derive an approximate expression for the test statistic of the Asimov data set using \cref{eq:multi-hypothesis-test-fc-test-statistic}
\begin{align}
    q_\text{A}
    &=
    -2 \sum_{i=1}^3
    \left(
    A^i
    \log
    \left(
    \frac{H_\text{BSM}^i}{H(0)^i}
    \right)
    \right)
    \\
    &=
    -2 \sum_{i=1}^3
    \left(
    H_\text{BSM}^i (1 + \delta^i)
    \log
    \left(
    \frac{H_\text{BSM}^i}{H_\text{BSM}^i (1 + \delta^i)}
    \right)
    \right)
    \\
    &\approx
    -2 \sum_{i=1}^3
    \left(
    - H_\text{BSM}^i (1 + \delta^i)   
    \left(
    \delta^i - \frac{(\delta^i)^2}{2}
    \right)
    \right)
    \\
    &\approx
    -2 \sum_{i=1}^3
    \left(
    - H_\text{BSM}^i 
   \frac{(\delta^i)^2}{2}
    \right)
    \\
    &=
    \sum_{i=1}^3
    \frac{(H(0)^i-H_\text{BSM}^i)^2}{H_\text{BSM}^i}
    \,,
\end{align}
where we have used the fact that the test statistic is maximised by $\mu=0$ for the Asimov data set (since $H(0) = H_\text{SM} = A$). We see that the test statistic of the Asimov data set can again be approximated by the {\bf Pearson's $\chi^2$ test statistic}
\begin{align}
    q_\text{A} \approx \chi^2_\text{A}
    &=
    \sum_{i=1}^3
    \frac{(H_\text{SM}^i-H_\text{BSM}^i)^2}{H_\text{BSM}^i}
    \\
    &=
    4.55
    \,.
\end{align}
We can then estimate the $p$-value for the hypothesis test using
\begin{align}
    p
    &=
    \int_{\chi^2_\text{A}}^\infty \frac{1}{2}f(q;1)\,dq
    \\
    &=
    \frac{1}{2}
   \text{erfc}\left(\sqrt{\frac{\chi^2_\text{A}}{2}}\right)
    \\
    &=
    1-
    \Phi\left(\sqrt{\chi^2_\text{A}}\right)
    \\
    &\approx
    0.016
    \,,
\end{align}
where erfc is the complementary error function and $\Phi$ is the cumulative distribution function of the normal distribution with mean 0 and standard deviation 1. This $p$-value differs somewhat from the exact $p$-value because the approximations used are not sufficiently accurate for the specific hypotheses in \cref{eq:exMultiDistSM} and \cref{eq:exMultiDistBSM}.

\subsection{Realistic Example: Hawking Radiation from a Black Hole Explosion}
\label{sec:realistic-example-m}

To illustrate the multinomial case, we consider what a future observation of a single black hole explosion in the vicinity of Earth could tell us about the particles present in nature, as discussed in Refs.~\cite{Baker:2021btk,Baker:2022rkn,Baker:2025zxm,Baker:2025ffi}.  As discussed in Ref.~\cite{Baker:2025zxm}, it is plausible to think that a current or upcoming gamma ray observatory, like HAWC~\cite{HAWC} or LHAASO~\cite{LHAASO}, could observe such an explosion.  The relevant observable is the time-dependent photon spectrum emitted by the black hole, which can be predicted under the SM hypothesis and within a specified BSM scenario. As a benchmark BSM model, we supplement the Standard Model particle content by an additional dark sector. This sector need not contain a dark matter candidate. It rather serves as a proxy for new degrees of freedom that couple only very weakly to SM fields. Since the coupling to the black hole is purely gravitational, the resulting emission depends only on the particle content and masses, and is insensitive to gauge or Yukawa interactions between the dark sector and the SM. For simplicity, we model the dark sector as $n_\chi$ Dirac fermions with a common mass $\Lambda_{\text{DS}}$, denoted $\mathrm{DS}(n_\chi,\Lambda_{\text{DS}})$.

In the BSM scenario, the black hole radiates dark sector particles in addition to SM particles, thereby reducing the fraction of the energy carried by SM particles and, in particular, the expected number of photons.\footnote{Note that this assumes that the BSM sector does not produce an appreciable number of secondary photons, see \cite{Baker:2022rkn}.} However, since the distance to the explosion can not be independently measured, the number of photons could be reduced either due to additional BSM degrees of freedom or simply from a larger source distance. The total photon count therefore does not provide any discriminating power. We instead normalize the photon spectrum to a fixed total count $N_\gamma$ and use only its shape information to distinguish the SM and BSM hypotheses.  In this case, the expected distribution will be described by a multinomial distribution (rather than Poisson as is typical in collider experiments).

\begin{figure}
\centering
\includegraphics[width=0.55\linewidth]{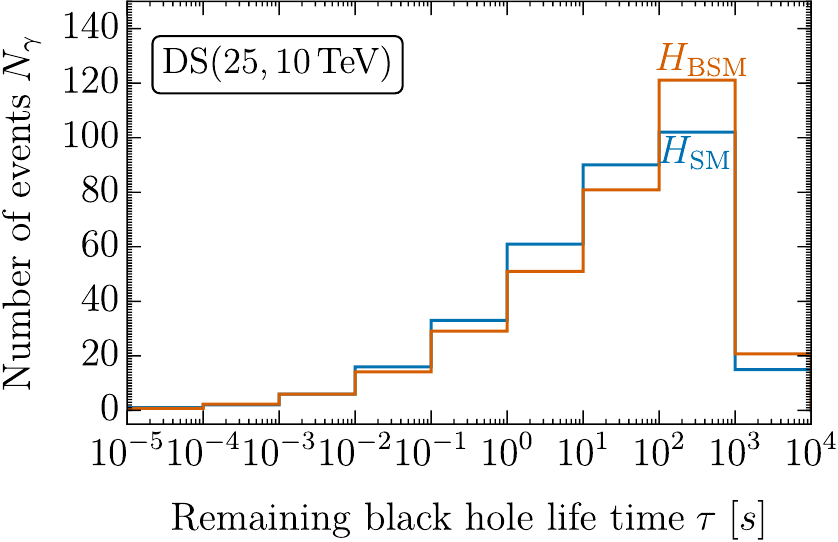}
\caption{Expected photon counts from an exploding Schwarzschild black hole at $d=0.02$pc under the SM hypothesis (blue) and photon signal under a BSM scenario with 25 new Dirac particles with a mass of 10 TeV (orange). The BSM photon count is normalized to $N_\gamma = 326$. The horizontal axis indicates the remaining lifetime of the black hole in seconds. We assume the detection is made by the HAWC telescope in its first zenith band and use $E_\text{min}=10^3$\,GeV.}
\label{fig:SM_observation}
\end{figure}

Taking neutrinos to be Majorana fermions, the expected photon counts in time bins equally spaced in log space from $10^{-5}$\,s to $10^4$\,s (so called `decadal binning') under the SM hypothesis for a black hole at $d_\text{SM}=0.02$\,pc and for the BSM benchmark point $\mathrm{DS}(25,10\,\mathrm{TeV})$ at $d_\text{BSM}=0.017$\,pc and detected in the central region of the HAWC telescope are
\begin{align}
H_\text{SM} 
&=
(0.853, 2.42, 6.40, 15.5, 33.2, 60.9, 89.9, 101, 15.4) \\
H_\text{BSM} 
&=
(0.840, 2.34, 6.02, 14.1, 29.1, 51.0, 80.9, 121, 20.7)
\,,
\end{align}
see Ref.~\cite{Baker:2025ffi} for details.  Both $H_\text{SM}$ and $H_\text{BSM}$ have a total photon count of $N_\gamma = 326$. The photon spectra are shown in \cref{fig:SM_observation} for the SM (blue) and the BSM hypothesis (orange). Since the photon count is fixed, the outcomes follow a multinomial distribution as discussed in \cref{sec:exact-p-value-m}.

Because no black hole explosion has yet been observed, we do not have real data. To make sensitivity projections we therefore adopt an Asimov data set, assuming the observed signal matches the Standard Model prediction for an explosion at a distance $d_\text{SM}=0.02\,\mathrm{pc}$,
\begin{align}
    A = H_\text{SM} 
    \,.
\end{align}
%
\begin{figure}
\centering
\includegraphics[width=0.49\textwidth]{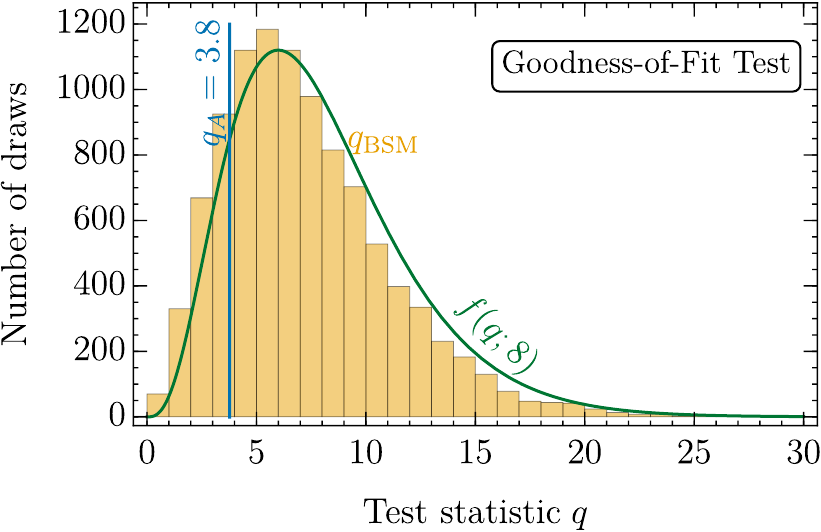}
\includegraphics[width=0.49\textwidth]{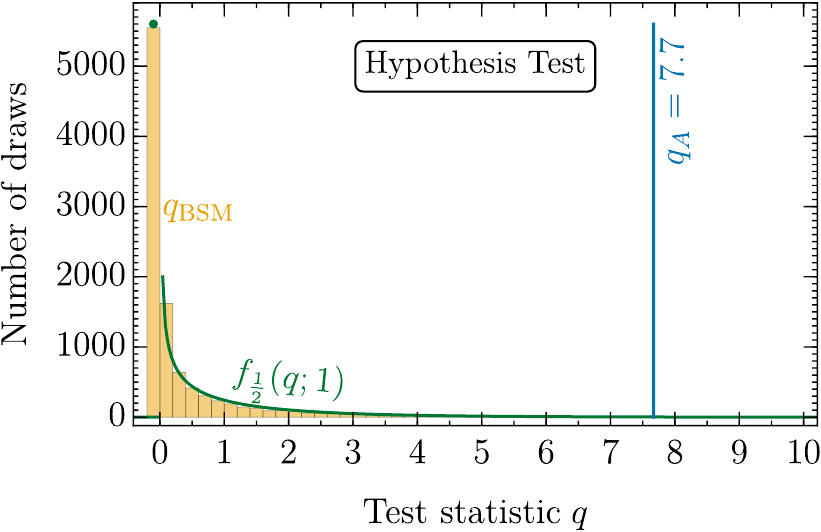}
\caption{The distribution of the test statistic obtained by Monte-Carlo sampling $H_\text{BSM}$ using a goodness-of-fit test (left) and a hypothesis test (right). The green lines show the asymptotic limits $f(q;8)$ (left) and $f_{\frac{1}{2}}(q;1)$ (right). The blue vertical lines show the test statistic of the Asimov data set, $q_\text{A}$.}
\label{fig:test-statistic}
\end{figure}
%
This situation was considered in Ref.~\cite{Baker:2025ffi} and a goodness-of-fit test was used to quantify the BSM model parameter space that could be excluded.  We now compare a goodness-of-fit analysis with a hypothesis test. Because the binning in this example is sufficiently fine that an exact enumeration of outcomes is impractical, we estimate the $p$-values using the Monte Carlo and asymptotic procedures described in \cref{subsec:M-MC-gof,subsec:M-MC-hypothesis-test}.

\Cref{fig:test-statistic} (left) shows the distribution of the goodness-of-fit test statistic, defined in \cref{eq:test-statistic-M-goodness-of-fit}, obtained from $10^4$ Monte Carlo draws generated assuming that $H_{\text{BSM}}$ is the true model of nature. The green curve shows the corresponding asymptotic description $f(q;k)$ from \cref{eq:p-value-M-goodness-of-fit-asymptotic} (with $f$ defined in \cref{eq:chi2-distribution}), i.e., a $\chi^2$ distribution with $k=8$ degrees of freedom.  We see that there is good (but not perfect) agreement.  In \cref{fig:test-statistic} (right) we show the distribution of the test statistic using the Wilks' theorem approach, defined in \cref{eq:test-statistic-M-hypothesis}, obtained from $10^4$ Monte Carlo draws generated assuming that $H_{\text{BSM}}$ is the true model of nature. We also show the corresponding asymptotic description $f_{\frac{1}{2}}(q;1)$ discussed in \cref{eq:test-statistic-M-hypothesis-asymptotic}, i.e., a $\tfrac{1}{2}\chi^2$ distribution with one degree of freedom. The blue vertical lines indicate the corresponding value of the test statistic for the Asimov data set.

\Cref{fig:test-statistic} provides a direct visual comparison of the expected $p$-values from a goodness-of-fit test and a hypothesis test. In each case, the $p$-value is given by the fraction of Monte Carlo draws with test statistic $q \ge q_A$. The goodness-of-fit distribution (left) contains a much larger tail beyond $q_A$ than the hypothesis-test distribution (right), leading to a correspondingly larger $p$-value. We again see that the hypothesis test will give a much smaller $p$-value than the goodness-of-fit test.

\begin{figure}
\centering
\includegraphics[width=0.49\textwidth]{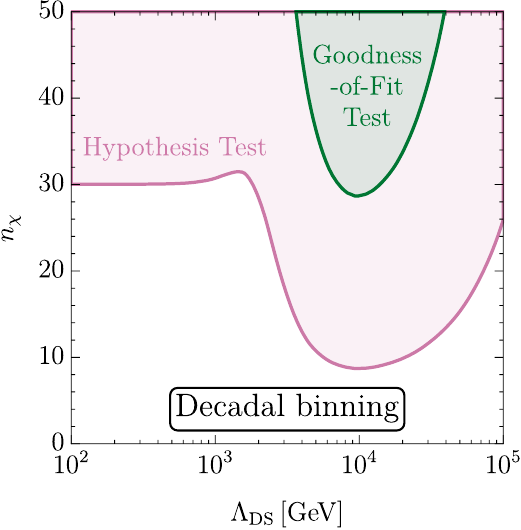}
\includegraphics[width=0.49\textwidth]{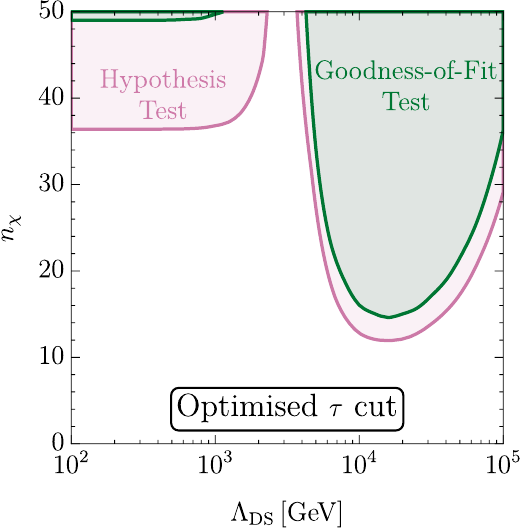}
\caption{Projected 95\% CL sensitivity of a goodness-of-fit test (green) and a hypothesis test (pink) for decadal binning (left, see \cref{fig:SM_observation}) and with an optimised $\tau$ cut (right) following \cite{Baker:2025ffi}.}
\label{fig:M-comparison}
\end{figure}

\Cref{fig:M-comparison} (left) shows the projected 95\% CL sensitivity of a goodness-of-fit test (green) and a hypothesis test (pink) in the BSM parameter space $(\Lambda_{\text{DS}}, n_\chi)$, using decadal binning (left) and two bins with an optimised $\tau_\text{cut}$ (right), with $E_\text{min} = 10^{3.1}$\,GeV (see Ref.~\cite{Baker:2025ffi} for details). Each curve is evaluated by computing, on a grid of parameter points, the Asimov value $q_A$ for the appropriate test statistic (namely, \cref{eq:test-statistic-asimov-M-goodness-of-fit} for goodness-of-fit testing, and \cref{eq:multi-hypothesis-test-fc-test-statistic} for hypothesis testing) and identifying the regions where $q_A$ exceeds the threshold values of the test statistics that correspond to $p=0.05$ (that is $q_A > 18.3$ for the goodness-of-fit test with 10 degrees of freedom, and $q_A > 2.71$ for the hypothesis test).\footnote{Our results in \cref{fig:M-comparison}, derived for $d_\text{SM} = 0.02\,$pc, are in agreement with fig.~8 of \cite{Baker:2025ffi}, which corresponds to  $d_\text{SM} = 0.01\,$pc.}  We see in both panels that the hypothesis test excludes a substantially larger region of parameter space than the goodness-of-fit test. In particular, whereas the goodness-of-fit test differentiates relatively heavy dark sectors, $\Lambda_{\text{DS}}\sim 10^{4.5}\,\text{GeV}$, the hypothesis test remains sensitive down to much smaller dark sector masses\footnote{Although we do not show it, the reach extends to $\Lambda_{\text{DS}}\to 0$.} and to scenarios with roughly $\mathcal{O}(20)$ fewer Dirac fermions.  We see that when comparisons are made with other sensitivity projections or with actual data, it is essential that the correct approach is taken.  In line with collider physics, we advocate that a hypothesis test should be used.

\Cref{fig:M-comparison} (right) shows the projected 95\% CL sensitivity of a goodness-of-fit test (green) and a hypothesis test (pink) for an optimized time-binning strategy discussed in Ref.~\cite{Baker:2025ffi}.  There it was shown that for goodness-of-fit tests the sensitivity can be improved by merging the decadal time bins into two bins separated at an optimized split point $\tau_{\text{cut}}^{\text{opt}}$, which depends on the BSM parameter point and on the minimum photon energy $E_{\min}$. Comparing the goodness-of-fit test between the left and right panels, we see that the sensitivity can always be improved through this method.  However, for the hypothesis test and for dark sector masses above $10^{3.5}$\,GeV, not only is there is no improvement but the optimised 2 bin provides less sensitivity than decadal binning.  With a hypothesis test and when statistical errors dominate, the sensitivity increases as the number of bins increases, until the main differences between the hypotheses are resolved, at which point finer binning does not dramatically increase (or decrease) the sensitivity.  This is in contrast to a goodness-of-fit test where the sensitivity may initially increase as more differences are resolved, but will decrease as the number of bins increases due to an increased number of statistical degrees of freedom in the test.

Finally, since Ref.~\cite{Baker:2025ffi} employed a goodness-of-fit test to obtain its sensitivity projections but we now advocate for a hypothesis test, we present the main results from that work using a hypothesis test.  We note that while the final result of the statistical procedures in Ref.~\cite{Baker:2025ffi} is consistent with our discussion here, we now believe eq.~(4.5) and (4.8) in Ref.~\cite{Baker:2025ffi} are incorrect and the test statistic should instead match the goodness-of-fit expressions given in \cref{eq:test-statistic-M-goodness-of-fit} and \cref{eq:test-statistic-asimov-M-goodness-of-fit}.

\begin{figure}
\centering
\includegraphics[width=0.49\textwidth]{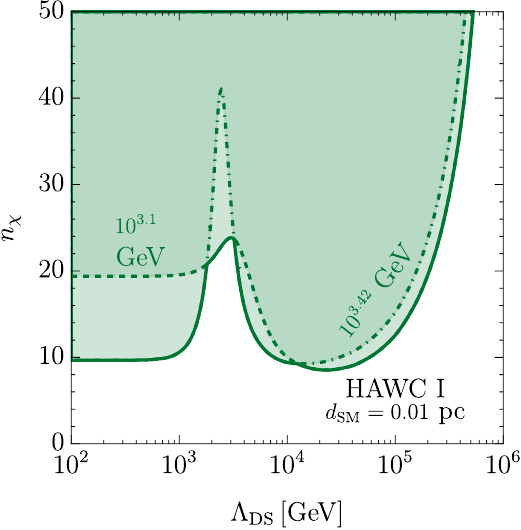}
\includegraphics[width=0.49\textwidth]{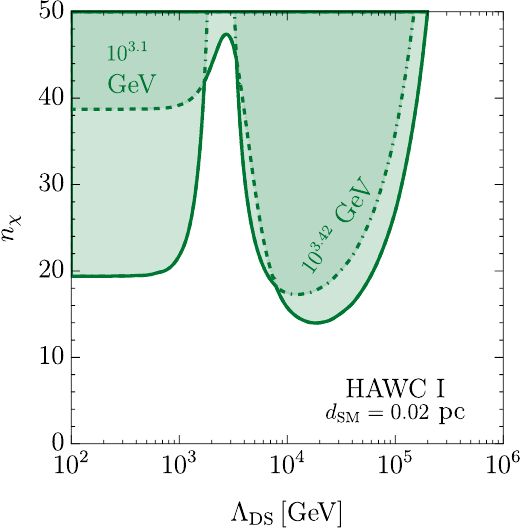}
\caption{The dark sector parameter space that can be probed by HAWC using a hypothesis test for different $E_\text{min}$ cuts, assuming the number of photons received under the SM hypothesis at 0.01 pc (left) and 0.02 pc (right).  This figure should be compared with fig.~10 in Ref.~\cite{Baker:2025ffi}, where a goodness-of-fit test was used.}
\label{fig:M-comparison-HAWC}
\end{figure}

In \cref{fig:M-comparison-HAWC} we show analogous results to fig.~10 in Ref.~\cite{Baker:2025ffi}, but using a hypothesis test and decadal binning with a Pearson's $\chi^2$ test approximation.  For each bin we assume $10\%$, uncorrelated systematic uncertainties, which we implement by computing the Asimov test statistic defined in \cref{eq:Pearson_ChiSquared_Systematics} with 10\% systematic errors, $\sigma_\text{sys}^i = 0.1 \,H^i_\text{BSM}$.  We show  $E_\text{min} = 10^{3.42}$\,GeV (dot-dashed), which gives better sensitivity at low masses, and $E_\text{min} = 10^{3.1}$\,GeV (dashed), which gives slightly better sensitivity at high masses and fills in the loss of sensitivity at $\Lambda_\text{DS} \approx 10^{3.5}$\,GeV.  On the left we show the results for the number of photons that HAWC would record for an explosion at $d_\text{SM} = 0.01$\,pc assuming only the Standard Model. In this case HAWC is expected to observe 413 photons with $E_\text{min} = 10^{3.42}$ and 1060 photons with $E_\text{min} = 10^{3.1}$\,GeV.  Comparing to figure 10 in Ref.~\cite{Baker:2025ffi}, we see that the exclusion region is similar at low  dark sector masses, but that a larger region is excluded at higher  masses.  On the right we show the same plot for a distance $d_\text{SM} = 0.02$\,pc.  We see that even at this larger distance, a significant portion of the parameter space we show could still be excluded.

\begin{figure}
\centering
\includegraphics[width=0.49\textwidth]{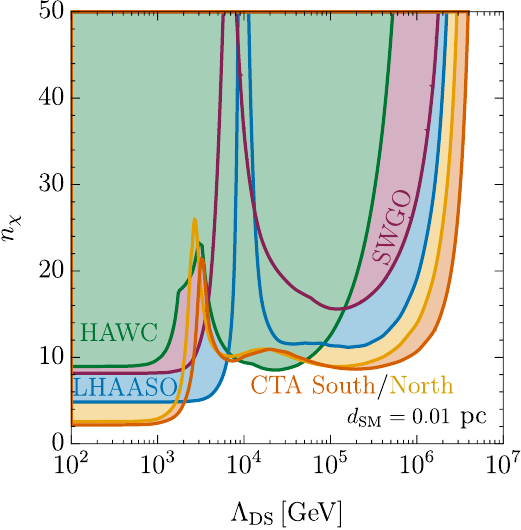}
\includegraphics[width=0.49\textwidth]{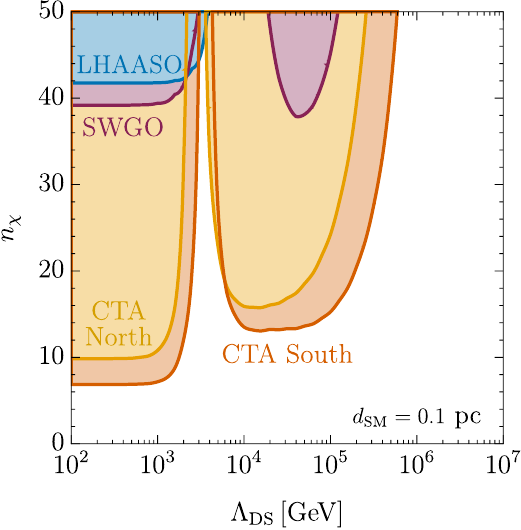}
\caption{Projected 95\% exclusion regions of the dark sector parameter space using a hypothesis test for explosions above HAWC,
LHAASO, SWGO and CTA North and South, in the optimistic scenarios that the black hole in each scenario spend as much time as needed in their fields of view.  This should be compared with fig.~12 in Ref.~\cite{Baker:2025ffi}, where a goodness-of-fit test was used.}
\label{fig:M-comparison-summary}
\end{figure}

In \cref{fig:M-comparison-summary} we show the 95\% exclusion regions similar to those in Figure 12 in Ref.~\cite{Baker:2025ffi}, but using a hypothesis test, decadal binning, and uncorrelated $10\%$ systematic uncertainties in each bin.  Comparing with the figures in that paper, we see that while the goodness-of-fit and hypothesis test contours are similar at the closer distance of $d_\text{SM} = 0.01$\,pc (when there are more photons detected), at the larger distance of $d_\text{SM} = 0.1$\,pc the hypothesis test leads to larger excluded regions, particularly for SWGO and LHAASO at smaller mass scales, $\Lambda_\text{DS}$, and for all experiments at large mass scales.

\section{The CLs Method}
\label{sec:cls}

We frequently compare sensitivity projections from future experiments with existing experimental observations in the same parameter space. This comparison requires a unified statistical treatment of both projections and observed constraints. In high energy particle physics, the statistical treatment of experimental data is well established \cite{Lyons:1986em,Barlow:1989:GUS,Cowan:1998ji,James:2006zz,Gross:2018okg,ParticleDataGroup:2026mpi,Segura:2024srj,bini-notes} but the exact presentation may differ between different experiments. In collider physics, for example, it is common practice to report experimental results in terms of the CLs method, rather than the $p$-values we have been using so far. 

\begin{figure}
\centering
\includegraphics[width=0.5\linewidth]{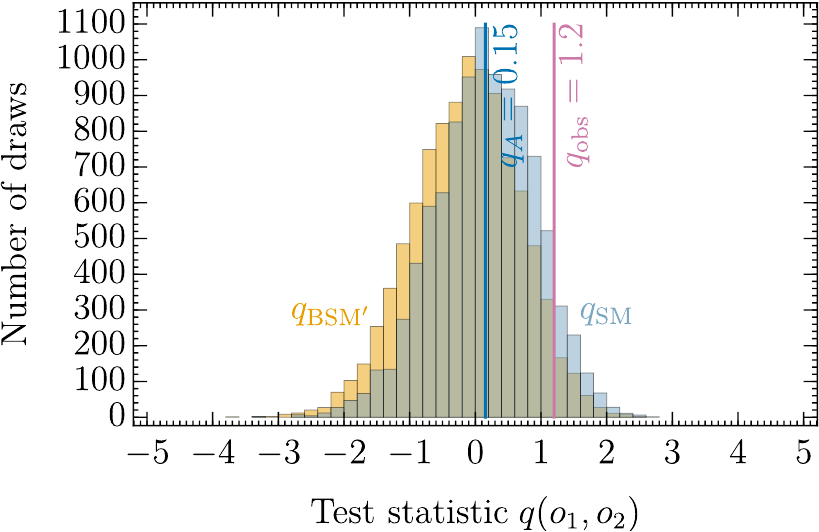}
\caption{Histograms of the hypothesis-test test statistics obtained from $10^4$ draws from $H'_\text{BSM}$ (orange) and from $H_\text{SM}$ (blue), along with the test statistic for the Asimov data set $q_\text{SM}$ (blue) and for some possible observation (pink).}
\label{fig:cls}
\end{figure}

In order to understand the motivation for the CLs method, proposed by Read~\cite{Read:2002hq} and building on the work of Birnbaum~\cite{Birnbaum1977}, we have to move away from projections and the Asimov data set and instead consider the case when an experiment has been run and an observation $o$, most likely not equal to $A$, has been found.   This observed data then leads to a test statistic, $q_\text{obs}$.  We imagine a situation where the experiment can not reliably distinguish $H'_\text{BSM}$ from $H_\text{SM}$, say
\begin{align}
    H_\text{SM} &= (11,14)
    \\
    H'_\text{BSM} &= (12,15) 
    \,,
\end{align}
and where the data is Poisson distributed. Using the test statistic for a hypothesis test and drawing Monte Carlo events in the way described in \cref{subsec:P-MC-hypothesis-test} leads to the distribution of test statistics shown in \cref{fig:cls}.  If we assume the Asimov data set with $q_A = 0.15$, we find a corresponding $p$-value of 
\begin{align}
    p_A &\approx \frac{3753}{10^4} 
    \\&= 0.38\,.
\end{align}
This implies that we cannot expect to rule out $H'_\text{BSM}$ at 95\% confidence level.  However, an actual observation may not agree exactly with the Asimov data set and may have, by chance, e.g., $q_\text{obs} = 1.2$.  This is not particularly unlikely if $H_\text{SM}$ is the true model of nature. Then we would find a $p$-value of
\begin{align}
    p_\text{obs} &\approx \frac{398}{10^4} 
    \\&= 0.04
\end{align}
and perhaps claim that $H'_\text{BSM}$ was ruled out at the 95\% confidence level. In fact, all that has happened is that the experiment can not decide between the two hypotheses but there has been a random fluctuation in the data away from $H'_\text{BSM}$. This reveals a limitation of quoting the $p$-value to decide which hypotheses to accept and reject.

The {\bf CLs method}~\cite{Read:2002hq} protects against these cases by also including the \textbf{power} of the experiment and observation.  The power is directly related to the probability of making a Type II error, denoted $\beta$. The power is then defined by $1-\beta$.  When using Monte Carlo methods, the power is calculated by counting the number of events drawn from the SM distribution (blue region in \cref{fig:cls}) with a test statistic $q > q_\text{obs}$ and then dividing by the total number of Monte Carlo samples. For our example, we obtain
\begin{align}
    1-\beta &= 
    \frac{779}{10^4}
    \\
    &\approx
    0.08
    \,.
\end{align}
When experimental data are unavailable, we use the Asimov dataset, setting $q_\text{obs} = q_\text{A}$. For the Asimov data set, we would expect $1-\beta = 0.5$.  

The CLs value is defined as
\begin{align}
    \text{CLs} &=
    \frac{\alpha}{1-\beta}
    \,,
\end{align}
where $\alpha$ corresponds to the $p$-value (or the probability of making a type I error). For $q_\text{obs}$ this gives
\begin{align}
    \text{CLs}_\text{obs} 
    &=
    0.51
    \,,
\end{align}
while for $q_\text{A}$ the value is similar with $\text{CLs}_\text{A}  = 0.74$. In experimental collider physics, CLs $< 0.05$ is typically shown as the excluded region.  We see that in this case the point in parameter space with $H'_\text{BSM}$ would not be excluded, which is reasonable as the experiment should not be able to decide between these two hypotheses and the small $p$-value only appeared to due to a fluctuation.  

We also see that when assuming the Asimov data set, the CLs value is twice the $p$-value, so CLs $< 0.05$ corresponds to $p < 0.025$. Note that this is not the same as the conventional significance threshold of the $p$-value, $p < 0.05$. We also note that while the CLs value is not strictly a probability, it has many features which make it well suited for determining exclusion limits.

\section{Converting a $p$-value to a Significance -- One-Sided vs Two-Sided Tests}
\label{sec:significance}

While here we have so far focused on $p$-values, in particle physics the $p$-value is often converted into a \textbf{significance}, $Z$.   The significance is typically quoted as the number of standard deviations such that the integral of a Gaussian distribution beyond $Z$ is equal to the $p$-value. The probability density function of a Gaussian distribution with mean $\mu$ and standard deviation $\sigma$ is
\begin{align}
    f(x)
    =
    \frac{1}{\sqrt{2\pi\sigma^2}}\exp
    \left(
    -\frac{(x-\mu)^2}{2\sigma^2}
    \right)
    \,.
\end{align}

\begin{figure}
\centering
\includegraphics[width=0.45\linewidth]{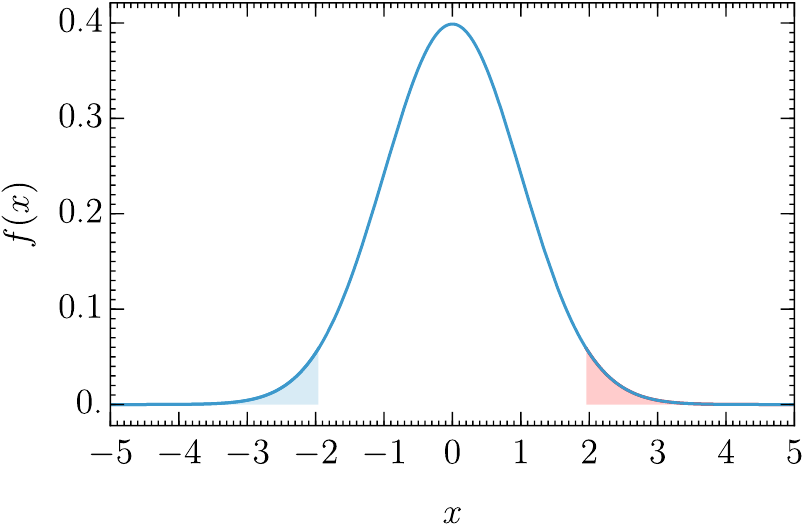}
\caption{A Gaussian Distribution with mean $\mu=0$ and standard deviation $\sigma=1$.  The blue and red regions have an area of 0.05 while the red region alone has an area of 0.025.}
\label{fig:Gaussian}
\end{figure}

One issue that can cause confusion is whether a two- or one-sided definition should be used.  That is, whether one should equate
\begin{align}
    p
    &=
    \int_{-\infty}^{\mu-Z} f(x)\,dx + \int_{\mu+Z}^\infty f(x)\,dx
    \,,
\end{align}
or
\begin{align}
    p
    &=\int_{\mu+Z}^\infty f(x)\,dx
    \,,
\end{align}
respectively.  In other words, should both the red and blue regions in \cref{fig:Gaussian} be counted, or only one of them.

\textbf{Goodness-of-Fit} tests are insensitive to whether the counts in a bin have more or less events than the BSM hypothesis. 
We can see this in \cref{fig:exactPValuePoissonGoodnessofFitPlot} where all possibilities outside the white contour are counted.  Due to this, a two-sided test should be used.  For a $p$-value of 0.05 we find $Z = 1.96 \sigma$, or a significance of $1.96\,\sigma$, which is often approximated as $2\,\sigma$.

A \textbf{Hypothesis Test}, on the other hand, depends on whether the number of counts in a bin is larger or smaller than in the BSM hypothesis.  For example, in \cref{fig:exactPValuePoissonHypothesisTestPlot} we only count the possibilities to the lower left of the black line.  We therefore use a one-sided test and now find $Z = 1.96$ for a $p$-value of 0.025.

We also note that for a hypothesis test, the significance is asymptotically given by $Z = \sqrt{q_\text{SM}} \sigma$.  We can see this since 
\begin{align}
    p &= \int_{\mu+Z}^\infty 
    \frac{1}{\sqrt{2\pi}\sigma}\exp
    \left(
    -\frac{(x-\mu)^2}{2\sigma^2}
    \right)\,dx
    \\
    &= \int_{Z}^\infty 
    \frac{1}{\sqrt{2\pi}\sigma}
    e^{
    -\frac{t^2}{2\sigma^2}}
    \,dt
    \,.
\end{align}
and as in \cref{eq:test-statistic-P-hypothesis-1-2-chi^2},
\begin{align}
    p
    &=
    \int_{q_\text{SM}}^\infty f_{\frac{1}{2}}(q;1)\,dq
    \\
    &=
    \int_{q_\text{SM}}^\infty \frac{1}{2}f(q;1)\,dq
    \\
    &=
    \int_{q_\text{SM}}^\infty \frac{1}{2}
    \frac{e^{-\frac{q}{2}}}{\sqrt{2\pi q}}
    \,dq
    \\
    &=
    \int_{\sqrt{q_\text{SM}}\sigma}^\infty \frac{1}{\sqrt{2\pi}\sigma}
    e^{-\frac{t^2}{2\sigma^2}}
    \,dt
    \,,
\end{align}
where we have introduced $q = t^2/\sigma^2$, so that
\begin{align}
    Z = \sqrt{q_\text{SM}} \sigma
    \,.
\end{align}
This is often used in the setting where the SM hypothesis is labelled background, $H_\text{SM} = B$, and the BSM hypothesis is labelled signal plus background, $H_\text{BSM} = S+B$.  If the signal is much smaller than the background in all bins, $S^i \ll B^i$, then the condition in \cref{eq:hypothesis-test-small-difference-condition} is satisfied and 
\begin{align}
    q_\text{SM} \approx \chi^2_\text{SM}
    &=
    \sum_{i}
    \frac{(H_\text{BSM}^i-H_\text{SM}^i)^2}{H_\text{BSM}^i}
    \\
    &=
    \sum_{i}
    \frac{(S^i)^2}{S^i+B^i}
    \\
    &\approx
    \sum_{i}
    \frac{(S^i)^2}{B^i}
    \,,
\end{align}
so that if there is just one bin then
\begin{align}
    Z &= 
    \sqrt{q_\text{SM}} \sigma
    \\
    &\approx 
    \frac{S}{\sqrt{S+B}} \sigma
    \\
    &\approx 
    \frac{S}{\sqrt{B}} \sigma
    \,.
\end{align}
The condition $S/\sqrt{B} > 2$ is often used as an approximation for when a point in BSM parameter space could be probed at the $2\,\sigma$ level.

\section{Systematic Errors}
\label{sec:systematic-errors}

Systematic errors can easily be included if Pearson's $\chi^2$ test statistic is a good approximation to the Asimov test statistic. In that case the statistical and systematic errors in each bin ($\sigma_\text{stat}^i = \sqrt{H_\text{BSM}^i}$ and $\sigma_\text{sys}^i$, respectively) can be added in quadrature as
\begin{equation}
\label{eq:Pearson_ChiSquared_Systematics}
    q_\text{A} \approx \chi^2_\text{A}
    =
    \sum_{i=1}^{n_B}
    \frac{(H_\text{SM}^i-H_\text{BSM}^i)^2}{H_\text{BSM}^i + \left(\sigma_\text{sys}^i \right)^2}
    \,.
\end{equation}
Note that $\sigma_\text{sys}^i$ are absolute errors, so a relative systematic error of, e.g., 10\% would correspond to $\sigma_\text{sys}^i = 0.1 \,H_\text{BSM}^i$.

If the systematic errors are correlated, we need to construct the covariance matrix, $V$, and Pearson's $\chi^2$ test statistic is generalised to 
\begin{equation}
\label{eq:Pearson_ChiSquared_Correlated_Systematics}
    q_\text{A} \approx \chi^2_\text{A}
    =
    \sum_{i,j =1}^{n_B}
    (H_\text{SM}^i-H_\text{BSM}^i) V^{-1}_{ij}  (H_\text{SM}^j-H_\text{BSM}^j)
    \,.
\end{equation}
where the diagonal entries of $V$ are $(\sigma_\text{stat}^i)^2 + (\sigma_\text{sys}^i)^2$, while the off-diagonal elements consist of the correlated systematic errors only.

Note that experimental data is sometimes reported in terms of a correlation matrix.  A typical example is the set of oblique parameters in the electroweak fit \cite{Fischer:2026bka}, which usually already combines both statistical and systematic errors.  Let's imagine that the FCC-ee (after the $WW$ run) can measure the electroweak oblique parameters with a precision \cite{deBlas:2016ojx}
\begin{align}
    \sigma_S &= 6.4 \times 10^{-3}\\	 
    \sigma_T &= 7.9 \times 10^{-3}\\	
    \sigma_U &= 5.4 \times 10^{-3}
\end{align}
with the correlation matrix $\rho$ equal to the current measured correlation matrix~\cite{Fischer:2026bka},
\begin{align}
    \rho = 
    \left(
    \begin{array}{ccc}
    \rho_{SS} & \rho_{ST} & \rho_{SU}\\
    \rho_{ST} & \rho_{TT} & \rho_{TU}\\
    \rho_{SU} & \rho_{TU} & \rho_{UU}
    \end{array}
    \right)
    =
    \left(
    \begin{array}{ccc}
    1 & 0.92 & -0.67\\
    0.92 & 1 & -0.88\\
    -0.67 & -0.88 & 1
    \end{array}
    \right)
    \,.
\end{align}
If we were just comparing $S$, $T$ and $U$ in our statistical test (and not combining it with any other observables) then the covariance matrix would be
\begin{align}
    V &=
    \left(
    \begin{array}{ccc}
    \rho_{SS}\sigma_S^2 & \rho_{ST}\sigma_S\sigma_T & \rho_{SU}\sigma_S\sigma_U\\
    \rho_{ST}\sigma_S\sigma_T & \rho_{TT}\sigma_T^2 & \rho_{TU}\sigma_T\sigma_U\\
    \rho_{SU}\sigma_S\sigma_U & \rho_{TU}\sigma_T\sigma_U & \rho_{UU}\sigma_U^2
    \end{array}
    \right)
\end{align}
which leads to
\begin{align}
    V^{-1} &=
    10^5
    \left(
    \begin{array}{ccc}
    3.6 & -4.3 & -2.7\\
    -4.3 & 5.8 & 4.1\\
    -2.7 & 4.1 & 3.5
    \end{array}
    \right)\,.
\end{align}
Since the Standard Model predicts $S = T = U = 0$ and assuming the BSM model predicts $S=H_\text{BSM}^S = 0.002$, $T=H_\text{BSM}^T = 0.002$ and $U=H_\text{BSM}^U = 0$, we would then have
\begin{align}
    q_A \approx \chi^2_A 
    &=
    \sum_{i,j\in \{S,T,U\}}
    H_\text{BSM}^i V^{-1}_{ij} H_\text{BSM}^j
    \\
    &=
    0.33
    \,.
\end{align}
Since the $\chi^2_A$ value is less than 2.71, this point would not be excluded at 95\% CL.  If another point is BSM parameter space predicts $S=H_\text{BSM}^S = 0.002$, $T=H_\text{BSM}^T = -0.002$ and $U=H_\text{BSM}^U = 0$, then we would have
\begin{align}
    q_A \approx \chi^2_A 
    &=
    7.2
    \,,
\end{align}
and this point would be excluded at 95\% CL.

If the systematic error is fully correlated, for instance a luminosity error in an invariant mass distribution at a collider, then the correlation matrix is
\begin{align}
    \rho
    &=
    \left(
    \begin{array}{ccccc}
    1 & 1 & 1 & \cdots & 1\\
    1 & 1 & 1 & \cdots & 1\\
    1 & 1 & 1 & \cdots & 1\\
    \vdots & \vdots & \vdots & \ddots & \vdots\\
    1 & 1 & 1 & \cdots & 1
    \end{array}
    \right)
\end{align}
and the systematic covariance matrix, $V_\text{sys}^{ij} = \rho^{ij}\sigma_\text{sys}^i\sigma_\text{sys}^j$, is
\begin{align}
    V_\text{sys}
    &=
    \left(
    \begin{array}{ccccc}
    (\sigma_\text{sys}^1)^2 & \sigma_\text{sys}^1\sigma_\text{sys}^2 & \sigma_\text{sys}^1\sigma_\text{sys}^3 & \cdots & \sigma_\text{sys}^1\sigma_\text{sys}^n\\
    \sigma_\text{sys}^2\sigma_\text{sys}^1 & (\sigma_\text{sys}^2)^2 & \sigma_\text{sys}^2\sigma_\text{sys}^3 & \cdots & \sigma_\text{sys}^2\sigma_\text{sys}^n\\
    \sigma_\text{sys}^3\sigma_\text{sys}^1 & \sigma_\text{sys}^3\sigma_\text{sys}^2 & (\sigma_\text{sys}^3)^2 & \cdots & \sigma_\text{sys}^3\sigma_\text{sys}^n\\
    \vdots & \vdots & \vdots & \ddots & \vdots\\
    \sigma_\text{sys}^n\sigma_\text{sys}^1 & \sigma_\text{sys}^n\sigma_\text{sys}^2 & \sigma_\text{sys}^n\sigma_\text{sys}^3 & \cdots & (\sigma_\text{sys}^n)^2
    \end{array}
    \right)
    \,.
\end{align}
In this case $V_\text{sys}$ may be singular (or badly conditioned) and does not have an inverse (or a reliable inverse), so a statistical error must be added as discussed above,
\begin{align}
    V = 
    \left(
    \begin{array}{ccccc}
    (\sigma_\text{stat}^1)^2 & 0 & 0 & \cdots & 0\\
    0 & (\sigma_\text{stat}^2)^2 & 0 & \cdots & 0\\
    \vdots & \vdots & \vdots & \ddots & \vdots\\
    0 & 0 & 0 & \cdots & (\sigma_\text{stat}^n)^2
    \end{array}
    \right)
    +
    V_\text{sys}
    \,.
\end{align}
If the systematic error is not correlated, then the correlation matrix is
\begin{align}
    \rho
    &=
    \left(
    \begin{array}{ccccc}
    1 & 0 & 0 & \cdots & 0\\
    0 & 1 & 0 & \cdots & 0\\
    0 & 0 & 1 & \cdots & 0\\
    \vdots & \vdots & \vdots & \ddots & \vdots\\
    0 & 0 & 0 & \cdots & 1
    \end{array}
    \right)
\end{align}
and \cref{eq:Pearson_ChiSquared_Correlated_Systematics} reduces to \cref{eq:Pearson_ChiSquared_Systematics}.

If Pearson's $\chi^2$ test statistic is not a good approximation to the Asimov test statistic, uncorrelated systematic errors can be incorporated as modifications to the likelihoods as outlined in Ref.~\cite{Conway:2011in} (see, e.g., Refs.~\cite{Acanfora:2026aay,Demortier:2007zz} for examples).

\section{Conclusions}
\label{sec:Conclusions}

In this work we have critically reviewed the use of goodness-of-fit tests and hypothesis tests in BSM phenomenology contexts.  We have shown how the exact $p$-values can be computed in both cases, and demonstrated both Monte Carlo methods and asymptotic limits for the distribution of test statistics and for the test statistic of the Asimov data set.  For hypothesis tests, we showed that a one-parameter family of alternate hypotheses leads to asymptotic limits.  We did this for both Poisson and multinomially distributed data, to demonstrate the same ideas in different contexts.  In all cases we saw that the test statistic of the Asimov data set can be approximated by Pearson's $\chi^2$ test statistic, and that the differences manifest themselves through the asymptotic distribution of tests statistics, which are typically $\chi^2$ or half-$\chi^2$ distributions with various degrees of freedom.  It is reasonable to think that the test statistic of the Asimov data set can be approximated by Pearson's $\chi^2$ test statistic in all cases due to the central limit theorem.  In the case of multinomially distributed data, we saw how the $G$-test arose as a better asymptotic approximation.

The main message that emerges from this discussion is that goodness-of-fit tests and hypothesis tests answer fundamentally different statistical questions, and give quantitively different results.  When projections are combined with existing experimental data, as is commonplace in BSM phenomenology, it is essential that the exclusion limits are answering the same question.  To this end, we advocate that hypothesis tests should always be used.  We hope we have provided enough resources within this document so that the average practitioner of BSM phenomenology can perform simple hypothesis tests with their models (in a variety of ways, if so desired).  We provide a \texttt{Mathematica} notebook with all the numerical calculations on \href{https://github.com/PhysicsBaker/Practical-Guide-to-Sensitivity-Projections}{github}.

We exemplify the quantitative differences between goodness-of-fit tests and hypothesis tests in two concrete scenarios, involving heavy resonances at a future muon collider and the future observation of an exploding black hole.  For the exploding black holes, we update results in the literature where a goodness-of-fit test was used and present the results for a hypothesis test.

We discuss the relationship between $p$-values and CLs values, and between $p$-values and significances.  While collider experiments typically use CLs values, any of these are reasonable as long as the method is clearly stated.  However, when exclusions are compared the methods used should agree.  We have also shown how asymptotically the condition $S/\sqrt{B} > 2$ is appropriate, as long as the signal is much smaller than the background.

Finally, we discuss systematic errors and demonstrate how to include uncorrelated and correlated errors in the test statistics of the Asimov data set.

\section*{Acknowledgements}

We would like to thank Christophe Grojean, Jesse Thaler and, in particular, Rafael Coelho Lopes de Sa for discussion in the early stages of this project. AT is grateful to Duccio Pappadopulo for generously sharing his expertise in statistics. MJB's work is partially supported by the US-NSF grant PHY-2609796. The work of AT is supported by the US-NSF grant PHY-2609766 and by an Alfred P.~Sloan research fellowship. FA, MJB and AT are grateful to the Aspen Center for Physics, which is supported by National Science Foundation grant PHY-2210452, where part of this work was performed. FA, MJB, JIJ and AT are also grateful to the Mainz Institute for Theoretical Physics (MITP) of the Cluster of Excellence PRISMA$^{+}$ (Project ID 390831469) for its hospitality and its partial support during a critical stage of this work.

\bibliographystyle{JHEP}
\bibliography{refs}

\providecommand{\href}[2]{#2}\begingroup\raggedright\begin{thebibliography}{10}

\bibitem{Lyons:1986em}
L.~Lyons, \emph{{STATISTICS FOR NUCLEAR AND PARTICLE PHYSICISTS}} (1986).

\bibitem{Barlow:1989:GUS}
R.J.~Barlow, \emph{Statistics: A Guide to the Use of Statistical Methods in the
  Physical Sciences}, Manchester Physics Series, Wiley-Blackwell, Chichester,
  England (1989).

\bibitem{Cowan:1998ji}
G.~Cowan, \emph{{Statistical data analysis}} (1998).

\bibitem{James:2006zz}
F.~James, \emph{{Statistical methods in experimental physics}} (2006).

\bibitem{Gross:2018okg}
E.~Gross, \emph{{Practical Statistics for High Energy Physics}},
  \href{https://doi.org/10.23730/CYRSP-2018-003.199}{\emph{CERN Yellow Rep.
  School Proc.} {\bfseries 3} (2018) 199}.

\bibitem{ParticleDataGroup:2026mpi}
{\scshape Particle Data Group} collaboration, \emph{{Review of Particle
  Physics}}, \href{https://doi.org/10.1142/s0217751x26300115}{\emph{Int. J.
  Mod. Phys. A} {\bfseries 41} (2026) 2630011}.

\bibitem{Segura:2024srj}
A.~Segura and A.C.~Parra, \emph{{A Practical Guide to Statistical Techniques in
  Particle Physics}},  [\href{https://arxiv.org/abs/2411.00706}{{\ttfamily
  2411.00706}}].

\bibitem{bini-notes}
C.~Bini, ``Data analysis in particle physics.''
  \url{https://www.roma1.infn.it/~bini/StatEPP_new.pdf}.

\bibitem{Moneta:2010pm}
L.~Moneta, K.~Belasco, K.S.~Cranmer, S.~Kreiss, A.~Lazzaro, D.~Piparo et~al.,
  \emph{{The RooStats Project}},
  \href{https://doi.org/10.22323/1.093.0057}{\emph{PoS} {\bfseries ACAT2010}
  (2010) 057} [\href{https://arxiv.org/abs/1009.1003}{{\ttfamily 1009.1003}}].

\bibitem{Abasov:2025hsy}
E.~Abasov, L.V.~Dudko, D.~Gorin and O.~Vasilevskii, \emph{{StatTestCalculator:
  A New General Tool for Statistical Analysis in High Energy Physics}},
  \href{https://doi.org/10.3103/S0027134925702765}{\emph{Moscow Univ. Phys.
  Bull.} {\bfseries 80} (2025) S913}
  [\href{https://arxiv.org/abs/2510.11637}{{\ttfamily 2510.11637}}].

\bibitem{Conway:2011in}
J.S.~Conway, \emph{{Incorporating Nuisance Parameters in Likelihoods for
  Multisource Spectra}},  in \emph{{PHYSTAT 2011}}, pp.~115--120, 2011,
  \href{https://doi.org/10.5170/CERN-2011-006.115}{DOI}
  [\href{https://arxiv.org/abs/1103.0354}{{\ttfamily 1103.0354}}].

\bibitem{cowan-slides}
G.~Cowan.
  \url{https://indico.global/event/6195/sessions/9305/attachments/25803/44567/cowan_santiago24_Lecture%203.pdf}.

\bibitem{Barlow:1993dm}
R.J.~Barlow and C.~Beeston, \emph{{Fitting using finite Monte Carlo samples}},
  \href{https://doi.org/10.1016/0010-4655(93)90005-W}{\emph{Comput. Phys.
  Commun.} {\bfseries 77} (1993) 219}.

\bibitem{Liu:2023bqc}
S.~Liu and C.~McGrew, \emph{{Binned Likelihood including Monte Carlo
  Statistical Uncertainty in Bayesian Inference}},
  [\href{https://arxiv.org/abs/2304.05433}{{\ttfamily 2304.05433}}].

\bibitem{Aylward}
A.~Aylward, D.J.~Fairbanks, M.~Kiladi and G.~Radick, \emph{Eugenics and the
  genetics society},
  \href{https://doi.org/10.1038/s41437-026-00838-5}{\emph{Heredity} {\bfseries
  135} (2026) 301}.

\bibitem{Resin:2023}
J.~Resin, \emph{{A Simple Algorithm for Exact Multinomial Tests}},
  \href{https://doi.org/10.1080/10618600.2022.2102026}{\emph{Journal of
  Computational and Graphical Statistics} {\bfseries 32} (2023) 539}.

\bibitem{Neyman:1933wgr}
J.~Neyman and E.S.~Pearson, \emph{{On the Problem of the Most Efficient Tests
  of Statistical Hypotheses}},
  \href{https://doi.org/10.1098/rsta.1933.0009}{\emph{Phil. Trans. Roy. Soc.
  Lond. A} {\bfseries 231} (1933) 289}.

\bibitem{Wilks:1938dza}
S.S.~Wilks, \emph{{The Large-Sample Distribution of the Likelihood Ratio for
  Testing Composite Hypotheses}},
  \href{https://doi.org/10.1214/aoms/1177732360}{\emph{Annals Math. Statist.}
  {\bfseries 9} (1938) 60}.

\bibitem{cowan-note}
G.~Cowan. \url{https://www.pp.rhul.ac.uk/~cowan/stat/notes/wilks.pdf}.

\bibitem{Baker:1983tu}
S.~Baker and R.D.~Cousins, \emph{{Clarification of the Use of Chi Square and
  Likelihood Functions in Fits to Histograms}},
  \href{https://doi.org/10.1016/0167-5087(84)90016-4}{\emph{Nucl. Instrum.
  Meth.} {\bfseries 221} (1984) 437}.

\bibitem{Cowan:2010js}
G.~Cowan, K.~Cranmer, E.~Gross and O.~Vitells, \emph{{Asymptotic formulae for
  likelihood-based tests of new physics}},
  \href{https://doi.org/10.1140/epjc/s10052-011-1554-0}{\emph{Eur. Phys. J. C}
  {\bfseries 71} (2011) 1554}
  [\href{https://arxiv.org/abs/1007.1727}{{\ttfamily 1007.1727}}], [Erratum:
  Eur.Phys.J.C 73, 2501 (2013)].

\bibitem{cowan-note-2}
G.~Cowan. \url{https://www.pp.rhul.ac.uk/~cowan/stat/medsig/medsigNote.pdf}.

\bibitem{Feldman:1997qc}
G.J.~Feldman and R.D.~Cousins, \emph{{A Unified approach to the classical
  statistical analysis of small signals}},
  \href{https://doi.org/10.1103/PhysRevD.57.3873}{\emph{Phys. Rev. D}
  {\bfseries 57} (1998) 3873}
  [\href{https://arxiv.org/abs/physics/9711021}{{\ttfamily physics/9711021}}].

\bibitem{Read:2002hq}
A.L.~Read, \emph{{Presentation of search results: The $CL_s$ technique}},
  \href{https://doi.org/10.1088/0954-3899/28/10/313}{\emph{J. Phys. G}
  {\bfseries 28} (2002) 2693}.

\bibitem{read-note}
A.L.Read.
  \url{https://www.ippp.dur.ac.uk/Workshops/02/statistics/proceedings//read.pdf}.

\bibitem{mike-evans-slides}
M.~Evans.
  \url{https://indico.cern.ch/event/1357458/attachments/2850539/4987501/Mike_Evans_Slides.pdf}.

\bibitem{read-slides}
A.L.Read.
  \url{https://indico.cern.ch/event/1357458/attachments/2850539/4988589/CLs_Physicist_Presentation.pdf}.

\bibitem{cowan-slides-2}
G.~Cowan.
  \url{https://www.pp.rhul.ac.uk/~cowan/stat/nikhef/cowan_nikhef_2011_4.pdf}.

\bibitem{gross-slides}
E.~Gross.
  \url{https://indico.cern.ch/event/687486/attachments/1638791/2615776/Statistics_CERN_2018_part1.pdf}.

\bibitem{bini-slides}
C.~Bini. \url{https://www.roma1.infn.it/~bini/StatEPP_new.pdf}.

\bibitem{sonneveld-slides}
J.~Sonneveld.
  \url{https://indico.cern.ch/event/637941/contributions/2606105/attachments/1510179/2358420/limits_statistics_201708_sonneveld.pdf}.

\bibitem{ATLAS:2011tau}
{\scshape ATLAS, CMS, LHC Higgs Combination Group} collaboration,
  \emph{{Procedure for the LHC Higgs boson search combination in Summer 2011}},
  .

\bibitem{cranmer-slides}
K.~Cranmer.
  \url{https://indico.cern.ch/event/126254/attachments/79215/113568/StatisticsLecture_CERN_2011_day1.pdf}.

\bibitem{vanderVaart1998}
A.W.~van~der Vaart, \emph{Asymptotic Statistics}, Cambridge University Press
  (1998).

\bibitem{ferguson2017course}
T.S.~Ferguson, \emph{A course in large sample theory}, Routledge (2017).

\bibitem{Kahlhoefer:2019vhz}
F.~Kahlhoefer, A.~M{\"u}ck, S.~Schulte and P.~Tunney, \emph{{Interference
  effects in dilepton resonance searches for $Z'$ bosons and dark matter
  mediators}}, \href{https://doi.org/10.1007/JHEP03(2020)104}{\emph{JHEP}
  {\bfseries 03} (2020) 104}
  [\href{https://arxiv.org/abs/1912.06374}{{\ttfamily 1912.06374}}].

\bibitem{Delahaye:2019omf}
J.P.~Delahaye, M.~Diemoz, K.~Long, B.~Mansouli{\'e}, N.~Pastrone, L.~Rivkin
  et~al., \emph{{Muon Colliders}},
  [\href{https://arxiv.org/abs/1901.06150}{{\ttfamily 1901.06150}}].

\bibitem{mu_smash}
H.~Al~Ali et~al., \emph{{The muon Smasher{\textquoteright}s guide}},
  \href{https://doi.org/10.1088/1361-6633/ac6678}{\emph{Rept. Prog. Phys.}
  {\bfseries 85} (2022) 084201}
  [\href{https://arxiv.org/abs/2103.14043}{{\ttfamily 2103.14043}}].

\bibitem{toward_muc}
C.~Accettura et~al., \emph{{Towards a muon collider}},
  \href{https://doi.org/10.1140/epjc/s10052-023-11889-x}{\emph{Eur. Phys. J. C}
  {\bfseries 83} (2023) 864}
  [\href{https://arxiv.org/abs/2303.08533}{{\ttfamily 2303.08533}}], [Erratum:
  Eur.Phys.J.C 84, 36 (2024)].

\bibitem{muon_accelerator_program}
R.B.~Palmer, \emph{Muon colliders},
  \href{https://doi.org/10.1142/S1793626814300072}{\emph{Reviews of Accelerator
  Science and Technology} {\bfseries 07} (2014) 137}
  [\href{https://arxiv.org/abs/https://doi.org/10.1142/S1793626814300072}{{\ttfamily
  https://doi.org/10.1142/S1793626814300072}}].

\bibitem{Black:2022cth}
K.M.~Black et~al., \emph{{Muon Collider Forum report}},
  \href{https://doi.org/10.1088/1748-0221/19/02/T02015}{\emph{JINST} {\bfseries
  19} (2024) T02015} [\href{https://arxiv.org/abs/2209.01318}{{\ttfamily
  2209.01318}}].

\bibitem{MuonCollider:2022xlm}
{\scshape Muon Collider} collaboration, \emph{{The physics case of a 3 TeV muon
  collider stage}},  [\href{https://arxiv.org/abs/2203.07261}{{\ttfamily
  2203.07261}}].

\bibitem{Narain:2022qud}
M.~Narain et~al., \emph{{The Future of US Particle Physics - The Snowmass 2021
  Energy Frontier Report}},
  [\href{https://arxiv.org/abs/2211.11084}{{\ttfamily 2211.11084}}].

\bibitem{P5:2023wyd}
{\scshape P5} collaboration, \emph{{Exploring the Quantum Universe: Pathways to
  Innovation and Discovery in Particle Physics}},
  [\href{https://arxiv.org/abs/2407.19176}{{\ttfamily 2407.19176}}].

\bibitem{InternationalMuonCollider:2024jyv}
{\scshape International Muon Collider} collaboration, \emph{{Interim report for
  the International Muon Collider Collaboration (IMCC)}},
  \href{https://doi.org/10.23731/CYRM-2024-002}{\emph{CERN Yellow Rep. Monogr.}
  {\bfseries 2/2024} (2024) 176}
  [\href{https://arxiv.org/abs/2407.12450}{{\ttfamily 2407.12450}}].

\bibitem{InternationalMuonCollider:2025sys}
{\scshape International Muon Collider} collaboration, \emph{{The Muon
  Collider}},  [\href{https://arxiv.org/abs/2504.21417}{{\ttfamily
  2504.21417}}].

\bibitem{Begel:2025ldu}
M.~Begel et~al., \emph{{United States Muon Collider Community White Paper for
  the European Strategy for Particle Physics Update}},
  [\href{https://arxiv.org/abs/2503.23695}{{\ttfamily 2503.23695}}].

\bibitem{Pappadopulo:2014qza}
D.~Pappadopulo, A.~Thamm, R.~Torre and A.~Wulzer, \emph{{Heavy Vector Triplets:
  Bridging Theory and Data}},
  \href{https://doi.org/10.1007/JHEP09(2014)060}{\emph{JHEP} {\bfseries 09}
  (2014) 060} [\href{https://arxiv.org/abs/1402.4431}{{\ttfamily 1402.4431}}].

\bibitem{Acanfora:2026aay}
F.~Acanfora, M.J.~Baker, T.~Martonhelyi and A.~Thamm, \emph{{Heavy Vector
  Triplets at a Muon Collider}},
  [\href{https://arxiv.org/abs/2605.13957}{{\ttfamily 2605.13957}}].

\bibitem{Baker:2021btk}
M.J.~Baker and A.~Thamm, \emph{{Probing the particle spectrum of nature with
  evaporating black holes}},
  \href{https://doi.org/10.21468/SciPostPhys.12.5.150}{\emph{SciPost Phys.}
  {\bfseries 12} (2022) 150}
  [\href{https://arxiv.org/abs/2105.10506}{{\ttfamily 2105.10506}}].

\bibitem{Baker:2025ffi}
M.J.~Baker, J.~Iguaz~Juan, A.~Symons and A.~Thamm, \emph{{Probing dark sectors
  with exploding black holes: gamma rays}},
  \href{https://doi.org/10.1007/JHEP06(2026)042}{\emph{JHEP} {\bfseries 06}
  (2026) 042} [\href{https://arxiv.org/abs/2512.19603}{{\ttfamily
  2512.19603}}].

\bibitem{Baker:2022rkn}
M.J.~Baker and A.~Thamm, \emph{{Black hole evaporation beyond the Standard
  Model of particle physics}},
  \href{https://doi.org/10.1007/JHEP01(2023)063}{\emph{JHEP} {\bfseries 01}
  (2023) 063} [\href{https://arxiv.org/abs/2210.02805}{{\ttfamily
  2210.02805}}].

\bibitem{Baker:2025zxm}
M.J.~Baker, J.~Iguaz~Juan, A.~Symons and A.~Thamm, \emph{{Could We Observe an
  Exploding Black Hole in the Near Future?}},
  \href{https://doi.org/10.1103/nwgd-g3zl}{\emph{Phys. Rev. Lett.} {\bfseries
  135} (2025) 111002} [\href{https://arxiv.org/abs/2503.10755}{{\ttfamily
  2503.10755}}].

\bibitem{HAWC}
``{HAWC: The High-Altitude Water Cherenkov Observatory}.''
  \url{https://www.hawc-observatory.org/}.

\bibitem{LHAASO}
``{LHAASO: The Large High Altitude Air Shower Observatory}.''
  \url{http://english.ihep.cas.cn/lhaaso/}.

\bibitem{Birnbaum1977}
A.~Birnbaum, \emph{{The Neyman-Pearson theory as decision theory, and as
  inference theory; with a criticism of the Lindley-Savage argument for
  Bayesian theory.}}, \href{https://doi.org/10.1007/BF00485690}{\emph{Synthese}
  {\bfseries 36} (1977) 19}.

\bibitem{Fischer:2026bka}
{\scshape Gfitter Group} collaboration, \emph{{The Higgs boson through the lens
  of electroweak precision data}},
  [\href{https://arxiv.org/abs/2607.09861}{{\ttfamily 2607.09861}}].

\bibitem{deBlas:2016ojx}
J.~de~Blas, M.~Ciuchini, E.~Franco, S.~Mishima, M.~Pierini, L.~Reina et~al.,
  \emph{{Electroweak precision observables and Higgs-boson signal strengths in
  the Standard Model and beyond: present and future}},
  \href{https://doi.org/10.1007/JHEP12(2016)135}{\emph{JHEP} {\bfseries 12}
  (2016) 135} [\href{https://arxiv.org/abs/1608.01509}{{\ttfamily
  1608.01509}}].

\bibitem{Demortier:2007zz}
L.~Demortier, \emph{{P values and nuisance parameters}},  in \emph{{PHYSTAT-LHC
  Workshop on Statistical Issues for LHC Physics}}, pp.~23--33, 2007.

\end{thebibliography}\endgroup

\end{document}